\documentclass[journal]{IEEEtran}

\usepackage{amsmath}
\usepackage{amssymb}
\usepackage{amsthm}
\usepackage{graphicx}
\usepackage{subfigure}
\usepackage{multirow}
\usepackage{subfigure}
\usepackage{color}
\usepackage{threeparttable}
\usepackage{bm}
\usepackage{stfloats}
\usepackage{bigstrut}
\usepackage{booktabs}
\usepackage{soul}
\usepackage{cite}
\usepackage[
colorlinks,
linkcolor=red,
anchorcolor=green,
filecolor=red,
urlcolor=red,
citecolor=green]{hyperref}
\usepackage[numbers,sort&compress]{natbib}
\soulregister\cite7
\soulregister\ref7

\newcommand{\RNum}[1]{\uppercase\expandafter{\romannumeral #1\relax}}

\begin{document}
	
	\title{Deep Decomposition and Bilinear Pooling Network for Blind Night-Time Image Quality Evaluation}
	
	\author{Qiuping~Jiang,
		Jiawu~Xu,
		Yudong~Mao,
		Wei~Zhou,
		Xiongkuo~Min,
		Guangtao~Zhai
		\thanks{Q. Jiang, J. Xu, and Y. Mao are with the School of Information Science and Engineering, Ningbo University, Ningbo 315211, China (e-mail: jiangqiuping@nbu.edu.cn).
			
		W. Zhou is with the Department of Electrical and Computer Engineering, University of Waterloo, Waterloo, Canada (e-mail: wei.zhou@uwaterloo.ca).
			
		X. Min and G. Zhai are with the Institute of Image Communication and Information Processing, Shanghai Jiao Tong University, Shanghai 200240, China ({minxiongkuo, zhaiguangtao}@sjtu.edu.cn).
		}
	}
	
	\markboth{IEEE Transactions on Image Processing}
	{Shell \MakeLowercase{\textit{et al.}}: Bare Demo of IEEEtran.cls for IEEE Journals}
	
	\maketitle

	\begin{abstract}
		Blind image quality assessment (BIQA), which aims to accurately predict the image quality without any pristine reference information, has been extensively concerned in the past decades. Especially, with the help of deep neural networks, great progress has been achieved. However, it remains less investigated on BIQA for night-time images (NTIs) which usually suffers from complicated authentic distortions such as reduced visibility, low contrast, additive noises, and color distortions. These diverse authentic degradations particularly challenges the design of effective deep neural network for blind NTI quality evaluation (NTIQE). In this paper, we propose a novel deep decomposition and bilinear pooling network (DDB-Net) to better address this issue. The DDB-Net contains three modules, i.e., an image decomposition module, a feature encoding module, and a bilinear pooling module. The image decomposition module is inspired by the Retinex theory and involves decoupling the input NTI into an illumination layer component responsible for illumination information and a reflection layer component responsible for content information. Then, the feature encoding module involves learning feature representations of degradations that are rooted in the two decoupled components separately. Finally, by modeling illumination-related and content-related degradations as two-factor variations, the two feature sets are bilinearly pooled together to form a unified representation for quality prediction. The superiority of the proposed DDB-Net has been well validated by extensive experiments on several benchmark datasets. The source code will be made available soon.
	\end{abstract}


	\IEEEpeerreviewmaketitle

	\section{Introduction}
	\IEEEPARstart{D}{ue} to the poor lighting condition in night-time, the captured night-time images (NTIs) are usually perceived with poor visibility and low visual quality. Given that high-quality NTIs are crucial for consumer photography and practical applications such as automated driving systems, many NTI quality/visibility enhancement algorithms have been proposed. However, the research efforts on designing objective quality metrics that can automatically quantify the visual quality of NTIs and compare the performance of different NTI enhancement algorithms remain limited, which hereby hinders the development of this field. Generally, objective image quality assessment (IQA) methods can be roughly divided into three categories, i.e., full-reference (FR), no-reference (NR), and reduced-reference (RR) \cite{ZHAI2020Survey}. Among them, FR and RR IQA methods require full and partial reference information, respectively. However, for the NTIs we concerned, there is usually no available pristine image to provide any reference information. Therefore, NR-IQA is more valuable for NTIs in this regard.
	
	Early studies on NR-IQA mainly focus on specific distortion types, i.e., assuming that a particular distortion type is known and then specific distortion-related features are extracted to predict image quality \cite{JPEG1,JPEG2,Sharpness1,Sharpness2,Noise,MDM,JiangTII}. Obviously, such the specificity limits their applications like the real-world night-time scenario. Although the rapid advances in the IQA community during the last decade push to produce general-purpose blind IQA (BIQA) methods \cite{BLIINDS-II,BRISQUE,CurveletQA,DIIVINE,NRSL,NFERM,GM-LOG,GWH-GLBP,SSEQ,BIQME,ILNIQE,NIQE,BSD,ye2012Unsupervised,xu2016Blind,jiang2015Supervised,jiang2018Optimizing,wu2016Blind} that can simultaneously work with a number of distortion types, their efficacies are still limited to synthetic distortions. This is evident by the fact that they usually validate their performance on legacy synthetic distortion benchmark databases where the distorted images are simulated from pristine corpus in laboratory. As a result, the existing general-purpose BIQA methods still cannot work well with the authentically distorted images like the NTIs captured in the real-world night-time scenario. Recently, inspired by the success of deep neural networks in many image processing and computer vision tasks, great progresses have also been achieved on deep learning-based BIQA. However, it remains less investigated on deep learning-based BIQA for NTIs which usually suffer from complicated authentic distortions such as reduced visibility, low contrast, additive noises, invisible details, and color distortions. The diverse authentic degradations in NTIs pose great challenges to the design of highly effective end-to-end deep network architectures for blind NTI quality evaluation (NTIQE). 
	
	To evaluate the visual quality of NTIs, Xiang et al. \cite{BNBT} first established a dedicated large-scale natural NTI database (NNID), which contains $2,240$ NTIs with $448$ different image contents captured by three different photographic equipments in real-world scenarios along with their corresponding subjective quality scores (obtained by conducting human subjective experiments). Then, a NR quality metric called BNBT is proposed by considering both brightness and texture features. The experimental results on NNID database have demonstrated an acceptable performance of BNBT, i.e., the predicted quality scores by BNBT are consistent with ground truth subjective quality scores. Despite its effectiveness, BNBT requires elaborately-designed handcrafted features, which enlightens us to adopt an end-to-end data-driven method by taking the advantage of deep learning.
	
	However, designing tailored end-to-end deep neural networks for blind NTIQE is non-trivial due to the diverse authentic degradations. The main challenge is that heterogeneous distortions in NTIs make it difficult to learn a unified mapping from input NTI to quality score. An important observation is that the commonly-encountered distortions in NTIs can have impacts on either illumination perception or content perception. For example, the color distortion and additive noise are only influential  in content perception while the reduced visibility and low contrast are only influential  in illumination perception. Thus, it is intuitive to consider decomposing the input NTI into two independent components with each component accounting for illumination information and content information, respectively. Assisted by such a tailored image decomposition process, the degradation features related to illumination perception and content perception can be better learned and then fused to facilitate blind NTIQE. 
	\begin{figure*}[!t]
		\centering
		\includegraphics[width=\linewidth]{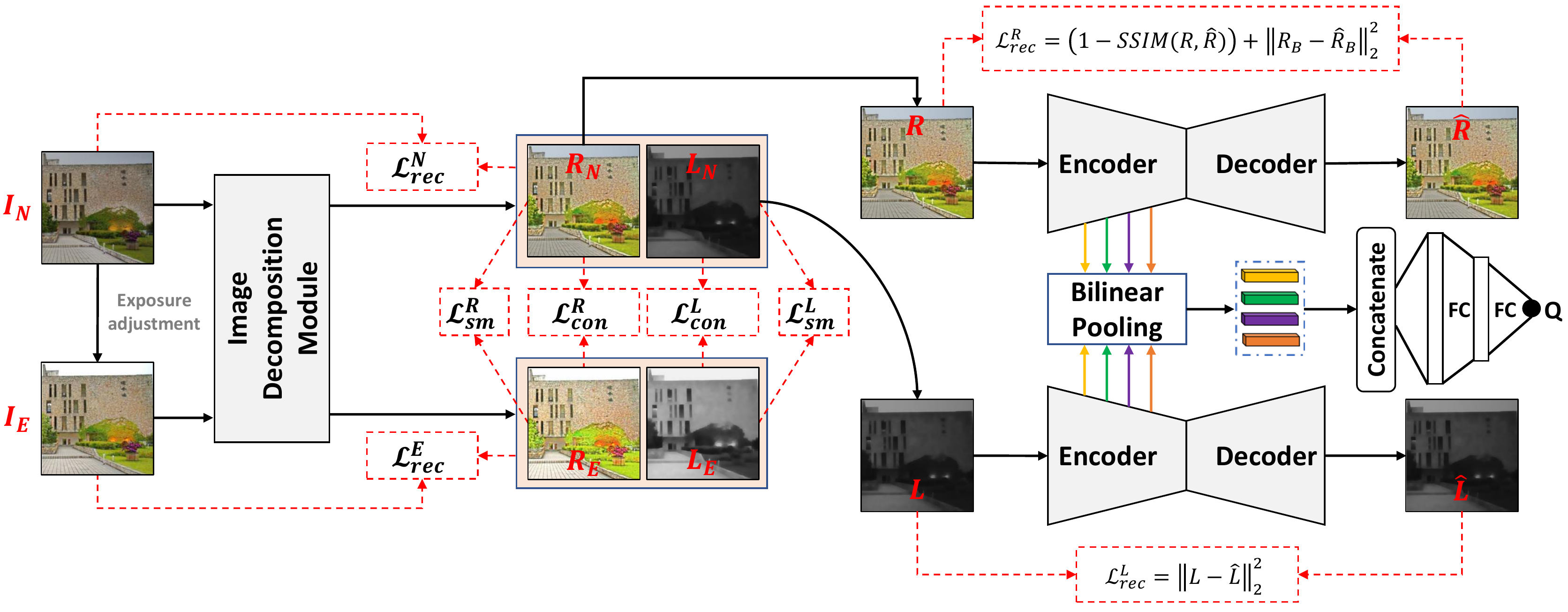}
		\caption{The proposed deep decomposition and bilinear pooling network (DDB-Net) for blind NTIQE. It contains an image decomposition module, a feature encoding module, and a bilinear pooling module. The image decomposition module takes an NTI as input and decouples it into two layer components, i.e., illumination ($L$) and reflection ($R$). Then, the feature encoding module involves learning feature representations of degradations that are rooted in the illumination and reflectance separately. Finally, the two feature sets are bilinearly pooled and concatenated together to form a unified representation for quality prediction.}
		\label{fig_DDBNet}
	\end{figure*}
	
	In this paper, we propose a novel deep decomposition and bilinear pooling network (DDB-Net) for blind NTIQE to better address the above issues. As shown in Fig. \ref{fig_DDBNet}, our DDB-Net contains three modules namely image decomposition module, feature encoding module, and bilinear pooling module. Inspired by the Retinex theory \cite{McCann2016}, the image decomposition module involves decoupling the input NTI into two layer components, i.e., one layer component (illumination) is responsible for illumination information, while the other one (reflection) for content information. Then, the feature encoding module involves learning feature representations of degradations that are rooted in the two decoupled components separately. Finally, by modeling illumination-related and content-related degradations as two-factor variations, the two feature sets are bilinearly pooled and concatenated together to form a unified representation for quality prediction. Extensive experiments conducted on several databases well demonstrated the superiority of the proposed DDB-Net against state-of-the-art BIQA methods. In summary, this paper presents the following contributions:
	
	1) We propose a novel ``decompose-and-conquer'' end-to-end deep neural network with the Retinex decomposition module embedding to better address the sophisticated blind NTIQE problem.
	
	2) We introduce a self-reconstruction-based feature encoding module and design tailored loss functions to regularize network training towards learning more targeted illunimation-related and content-related feature representations from the two decomposed components separately.
	
	3) We model the illumination-related and content-related degradations as two-factor variations and perform bilinear pooling to fuse the two sets of features into a unified feature representation for quality prediction of NTIs.
	
	
	The rest of this paper is organized in the following manner. Section II introduces the related works. Sections III illustrates the proposed method with details. Section IV presents the experimental results. Section V concludes the paper.
	
	\section{Related Works}
	In this section, we will review the existing related works, including traditional blind image quality assessment, deep learning-based blind image quality assessment, and blind image quality assessment in poor conditions.
	
	\subsection{Traditional Blind Image Quality Assessment}
	In the literature of traditional blind image quality assessment, natural scene statistics (NSS) and human visual system (HVS) are two main cues for designing objective BIQA models. As for NSS-based frameworks, Moorthy et al. \cite{DIIVINE} proposed the Distortion Identification-based Image Verity and INtegrity Evaluation (DIIVINE) index to evaluate perceptual image quality in a no-reference manner, which is composed of distortion identification and NSS-based quality regression. Likewise, the CurveletQA \cite{CurveletQA} also performs within a two-stage framework containing distortion classification and quality assessment. Different from DIIVINE, the quality assessment of CurveletQA is based on NSS features in curvelet domain. Except for the two-stage frameworks, other NSS-based BIQA algorithms have been developed. For example, Mittal et al. \cite{BRISQUE} presented the blind/referenceless image spatial quality evaluator (BRISQUE), which is operated in the spatial domain. Moreover, the BLIINDS-II \cite{BLIINDS-II} was proposed by exploiting the NSS model of discrete cosine transform (DCT) coefficients. In addition, some opinion-unaware BIQA methods based on NSS, i.e. so-called ``completely blind'' models, such as Natural Image Quality Evaluator (NIQE) \cite{NIQE} and ILNIQE \cite{ILNIQE}, have shown competitive performance with the help of massive natural images.
	
	Beyond the NSS features, many other statistical factors have been considered by researchers. In the family of two-stage framework, Liu et al. \cite{SSEQ} proposed the Spatial-Spectral Entropy-based Quality (SSEQ) index, where local spatial and spectral entropy features are used to predict perceptual image quality. The GM-LOG method \cite{GM-LOG} extracts the joint statistics of local contrast features to assess image quality, including gradient magnitude and Laplacian of Gaussian response. Apart from statistical structural features, lumninance histogram is used in the NRSL model \cite{NRSL}. The NSS features are combined with contrast, sharpness, brightness and colorfulness, together forming the BIQME framework \cite{BIQME}.
	
	For the HVS-based objective BIQA methods, some HVS-inspired features are applied to estimate perceptual image quality. Among these methods, Gu et al. \cite{NFERM} proposed the No-reference Free Energy-based Robust Metric (NFERM) on the basis of free energy principle. In \cite{BSD}, Li et al. used contrast masking to design the BIQA model based on structural degradation. Besides, according to the similar concept regarding the HVS properties, they proposed the GWH-GLBP by computing the gradient-weighted histogram of local binary pattern \cite{GWH-GLBP}.
	
	However, the above-mentioned conventional BIQA methods generally need to design elaborate handcrafted features with the pre-defined NSS or HVS mechanisms. Thus, resorting to data-driven methods based on deep learning is a promising alternative.
	
	\subsection{Deep Learning-based Blind Image Quality Assessment}
	Recently, deep learning has achieved great success in the field of blind image quality assessment. These methods can be typically divided into two categories, consisting of those using pre-trained deep features and end-to-end learning ones. For the first category, Wu et al. \cite{wu2017hierarchical} proposed the HFD-BIQA that integrates deep semantic features from ResNet \cite{he2016deep} into local structure features. Moreover, a Network in Network (NIN) model \cite{li2016no} pre-trained on ImageNet \cite{deng2009imagenet} was utilized to make image quality prediction. For the second category, Kang et al. \cite{kang2014convolutional} proposed a relatively shallow convolutional neural network (CNN) structure for BIQA. Each patch is assigned a subjective quality of the corresponding image as the ground-truth targets for training. In this way, the visual quality of whole image is then calculated by averaging predicted patch quality values. Furthermore, a BIQA model was developed based on shearlet transform and stacked auto-encoders \cite{li2015no}. In \cite{liu2017rankiqa}, the RankIQA was designed to synthesize masses of ranked images for training a Siamese network. Ma et al. \cite{ma2017end} proposed an end-to-end optimized deep neural Network for BIQA. Additionally, Bosse et al. \cite{bosse2017deep} presented the end-to-end WaDIQaM that can blindly learn perceptual image quality. The deep bilinear convolutional neural network called DBCNN was proposed to bilinearly pool the feature representations to a single quality score \cite{zhang2018blind}.
	
	Although the deep learning-based BIQA models can deliver good performance, they are not suitable for evaluating the perceptual quality of NTIs. This is mainly because these models usually neglect the specific characteristics of NTIs, e.g. reduced visibility, low contrast, additive noises, invisible details, and color distortions.
	
	\subsection{Quality Assessment for Images Captured in Poor Conditions}
	In real-world applications, people may encounter many kinds of poor imaging environments, e.g. hazy, rainy, underwater, and so on. In such poor conditions, capturing images with high-quality is quite challenging and thus addressing the blind quality assessment issue is urgently needed.
	
	In the quality evaluation of hazy images, Min et al. \cite{min2018objective} proposed the haze-removing features, structure-preserving features, and over-enhancement features to construct the objective quality assessment index. They also used synthetic hazy images to build an effective quality assessment model for image dehazing \cite{min2019quality}. For image deraining applications, a novel deep quality assessment model to predict the visual quality of real-world derained images was developed, which is achieved by designing a feature embedding network \cite{wu2020subjective}. As for underwater image quality evaluation, Yang et al. \cite{yang2015underwater} proposed to linearly combine chroma, saturation and contrast factors for quantifying the perceptual quality of underwater images. In \cite{guo2021underwater}, colorfulness, sharpness and contrast measures were fused to predict the underwater image quality. Most recently, Jiang et al. \cite{jiang2022underwater} first constructed a large-scale benchmark dataset for quality evaluation of underwater image enhancement and then proposed a no-reference underwater image quality metric by extracting both color and luminance features.
	
	It should be noted that the degradation characteristics of NTIs are different from those of hazy, rainy and underwater images. Thus, these quality evaluation methods developed for hazy, rainy and underwater images cannot achieve good performance on NTIs. In this paper, we focus on designing efficient end-to-end deep network architectures for blind NTIQE.
	
	\section{DDB-Net}
	In this section, we first describe the architecture of the image decomposition module. Then, we introduce the self-reconstruction-based feature encoding module for illunimation-related and content-related degradation feature learning. Finally, we introduce the bilinear pooling module for fusing the two sets of features.
	
	\begin{figure*}[!t]
		\centering
		\includegraphics[width=0.9\linewidth]{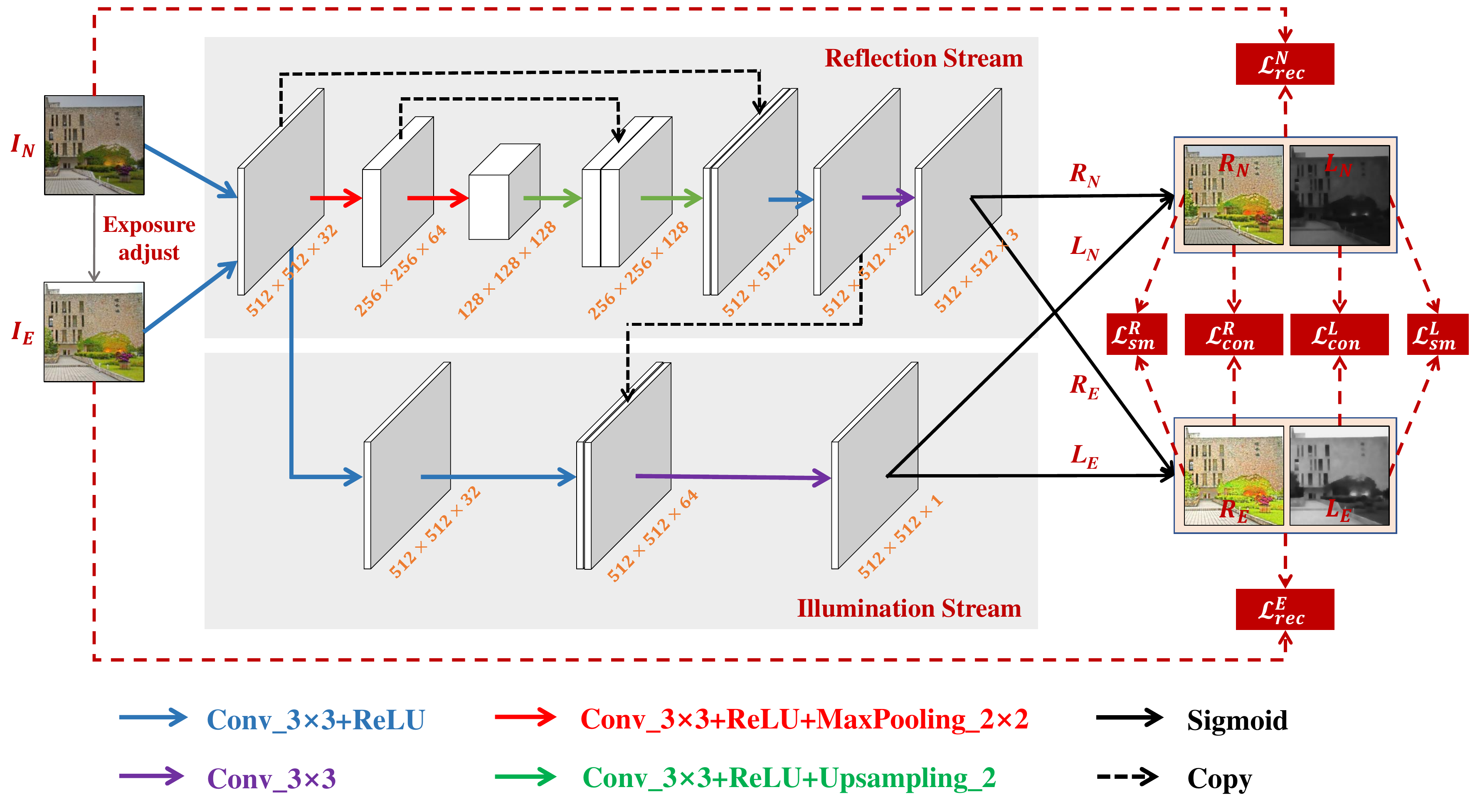}
		\caption{Detailed architecture of the image decomposition module. }
		\label{fig_ILD}
	\end{figure*}
	
	\subsection{Image Decomposition Module}
	According to the Retinex theory \cite{McCann2016}, a single image $I$ can be considered as a composition of two independent layer components, i.e., reflectance $R$ and illumination $L$, in the fashion of $I=R \otimes L$, where $\otimes$ denotes element-wise product. However, recovering two components from one single input is an ill-posed problem. Although the Retinex theory has been widely applied in relevant applications \cite{Liu2021BenchmarkingLI,Wang2019ProgressiveRM,Li2018StructureRevealingLI,Zhang2017FastHR}, there are obvious differences between our proposed method and these works. First, the problem we focuses is different from these works, i.e., we focus on the problem of NTIQE while these works \cite{Liu2021BenchmarkingLI,Wang2019ProgressiveRM,Li2018StructureRevealingLI,Zhang2017FastHR} focus on low-light/night-time image enhancement. Second, none of them have successfully decomposed a single image into two independent components with an embedded deep neural network module in a fully unsupervised manner. In what follows, we will present how to design an effective deep image decomposition module to achieve this goal. 
		
	The detailed architecture of our deep image decomposition module is shown in Fig. \ref{fig_ILD}. It contains two streams corresponding to the reflection component ($R$) and illumination component ($L$), respectively. The reflection stream adopts a 5-layer U-Net-like architecture, followed by two convolutional (conv) layers and a Sigmoid layer in the end, while the illumination stream is composed of two conv+ReLU layers and a conv layer on concatenated feature maps from the reflection stream, finally followed by a Sigmoid layer in the end. 
	
	Typically, there is no available ground truth reflection and illumination maps that can be used for supervised training. Therefore, designing a well-defined \textit{non-reference} loss function is the key to the success for training a stable deep image decomposition module. In the literature, a basic assumption is that different shots of the same scene should have the same reflection component. Furthermore, although the illumination maps could be intensively varied, they should be of simple and mutually consistent structures. These inspire us to take a pair of images (associated with the same scene) as input and impose both reflection and illumination constraints between the image pair to train the image decomposition module without ground truth.
	
	Specifically, during the training stage, the input to our image decomposition module is a pair of an NTI and its corresponding exposure-adjusted image (EAI). We denote the input NTI and EAI by $[I_N,I_E]$. Their corresponding reflection and illumination components are denoted by $[R_N,R_E]$ and $[L_N,L_E]$, respectively. The image decomposition module is constrained by a hybrid loss defined upon these components. In th following, we first describe how to generate the EAI from the input NTI, and then illustrate the definitions of different loss terms.
	
	\textit{EAI generation:} Given an input NTI $I_N$, we first generate multiple intermediate images with different exposure levels according to a camera response model that characterizes the relationship between pixel values and exposure ratios. Then, the multiple intermediate images with different exposure levels are fused to obtain an EAI $I_E$.
	
	Since there is no available camera information, an existing camera response model \cite{Ying2017ABM} that characterizes the general relationship between pixel value and exposure ratio is applied:
	\begin{equation}
		\mathcal{E}(I,e)=I^{(e^\alpha)} \cdot e^{\beta(1-e^\alpha)},
	\end{equation}
	where $I$ and $e$ represent the pixel value and the exposure ratio, respectively, and the parameters $\alpha=-0.3293$ and $\beta=1.1258$ are estimated by fitting a total number of 201 real-world camera response curves provided in the DoRF database \cite{Grossberg2004ModelingTS}. Specifically, the exposure ratios are $e,\cdots,e^K$, where the base ratio is empirically set to $e=2.4$ and the number of ratios is set to $K=4$, as in \cite{Liang2022RecurrentEG}. Based on the multi-exposure images $\{{\mathcal{E}}_1,{\mathcal{E}}_2,{\mathcal{E}}_3,{\mathcal{E}}_4\}$, the SPD-MEF algorithm \cite{SPD-MEF} is performed to reconstruct a fused image which can be used as the EAI $I_E$.
	
\begin{figure}[!t]
		\centering
		\includegraphics[width=0.85\linewidth]{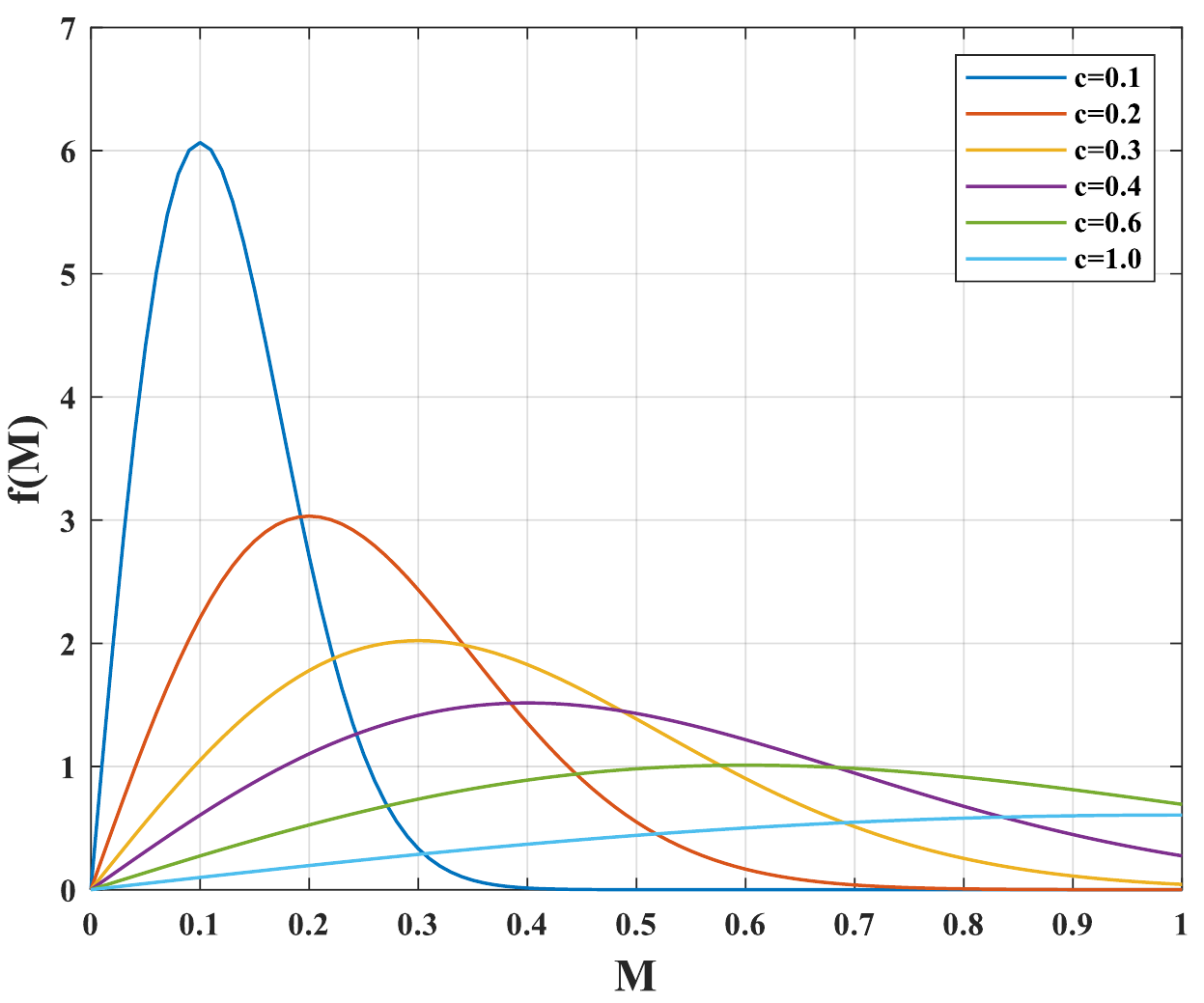}
		\caption{Penalty curves with different values of $c$.}
		\label{fig_funCurves}
	\end{figure}

	\textit{Inter-consistency loss:} The inter-consistency loss includes reflection consistency loss and illumination mutual consistency loss. First, the reflection consistency loss ${\mathcal{L}}_{con}^{R}$ encourages the reflection similarity, which is defined as follows:
	\begin{equation}
		{\mathcal{L}}_{con}^{R}= {\left\Vert {R}_{N}-{R}_{E} \right\Vert}_1,
	\end{equation}
	where ${\Vert \cdot \Vert}_1$ means the ${\ell}_1$ norm. Second, the illumination mutual consistency loss ${\mathcal{L}}_{con}^L$ is defined as follows:
	\begin{equation}
		{\mathcal{L}}_{con}^L=f(M)={\left\Vert \frac{M}{{c}^2} \otimes \text{exp}\left(-\frac{M^2}{2{c}^2}\right) \right\Vert}_1,
	\end{equation}
	\begin{equation}
		M=\left\vert \triangledown {L}_{N} \right\vert+\left\vert \triangledown {L}_{E} \right\vert,
	\end{equation}
	where $\triangledown$ means the first order derivative operator along both horizontal and vertical directions, $c$ is a parameter controlling the shape of the above penalty curve. To facilitate understanding, we draw the penalty curves with different values of $c$ in Fig. \ref{fig_funCurves}. As we can see, the penalty value first increases and then decreases to zero as $M$ increases. In our implementation, we set $c=0.1$. By minimizing such an illumination mutual consistency loss, the mutual strong edges are encouraged to be well preserved and all weak edges are to be suppressed. 
	
	
	\textit{Intra-smoothness loss:} Besides the inter-consistency loss mentioned above, we also consider intra-component smoothness loss. On the one hand, the illumination maps should be piece-wise smooth, thus we introduce a structure-aware smoothness loss ${\mathcal{L}}_{sm}^L$ to constraint both $L_N$ and $L_E$:
	\begin{equation}
		{\mathcal{L}}_{sm}^L={\left\Vert \frac{\triangledown L_N} {max\{(\triangledown R_N)^2,\tau\}}\right\Vert }_1 + {\left\Vert \frac{\triangledown L_E}{max\{(\triangledown {R}_{E})^2,\tau\}}\right\Vert }_1,
	\end{equation}
	where $\tau$ denotes a small positive constant which is empirically set to $\tau=0.01$ to avoid the denominator being zero. This loss measures the relative structure of the illumination with respect to the reflection. Therefore, the illumination loss can be aware of image structure which is reflected by reflection. Specifically, for a strong edge point in the reflection map, the penalty on the illumination will be small; for a point in the flat region of the reflection map, the penalty on the illumination becomes large. On the other hand, different from the illumination maps that should be piece-wise smooth, the reflectance maps are usually tend to be piece-wise continuous. Thus, we directly use a total-variation loss ${\mathcal{L}}_{sm}^R$ to constraint both $R_N$ and $R_E$:
	\begin{equation}
		{\mathcal{L}}_{sm}^R={\left\Vert \triangledown R_N \right\Vert}_1+{\left\Vert \triangledown R_E \right\Vert}_1.
	\end{equation}
	
	\textit{Reconstruction loss:} The third consideration is that the decomposed two components should well reproduce the input in the fashion of element-wise product, which is constrained by an image reconstruction loss:
	\begin{equation}
		{\mathcal{L}}_{rec}={\left\Vert I_N - L_N \otimes R_N \right\Vert}_1 + {\left\Vert I_E - L_E \otimes R_E \right\Vert}_1.
	\end{equation}
	
	\textit{Image decomposition loss:} The total loss for our image decomposition module is defined as follows:
	\begin{equation}
		{\mathcal{L}}_{idm}={\mathcal{L}}_{con}^R+{\mathcal{L}}_{con}^L+{\mathcal{L}}_{sm}^R+{\mathcal{L}}_{sm}^L+{\mathcal{L}}_{rec},
	\end{equation}
	
	\begin{figure*}[!t]
		\centering
		\includegraphics[width=0.9\linewidth]{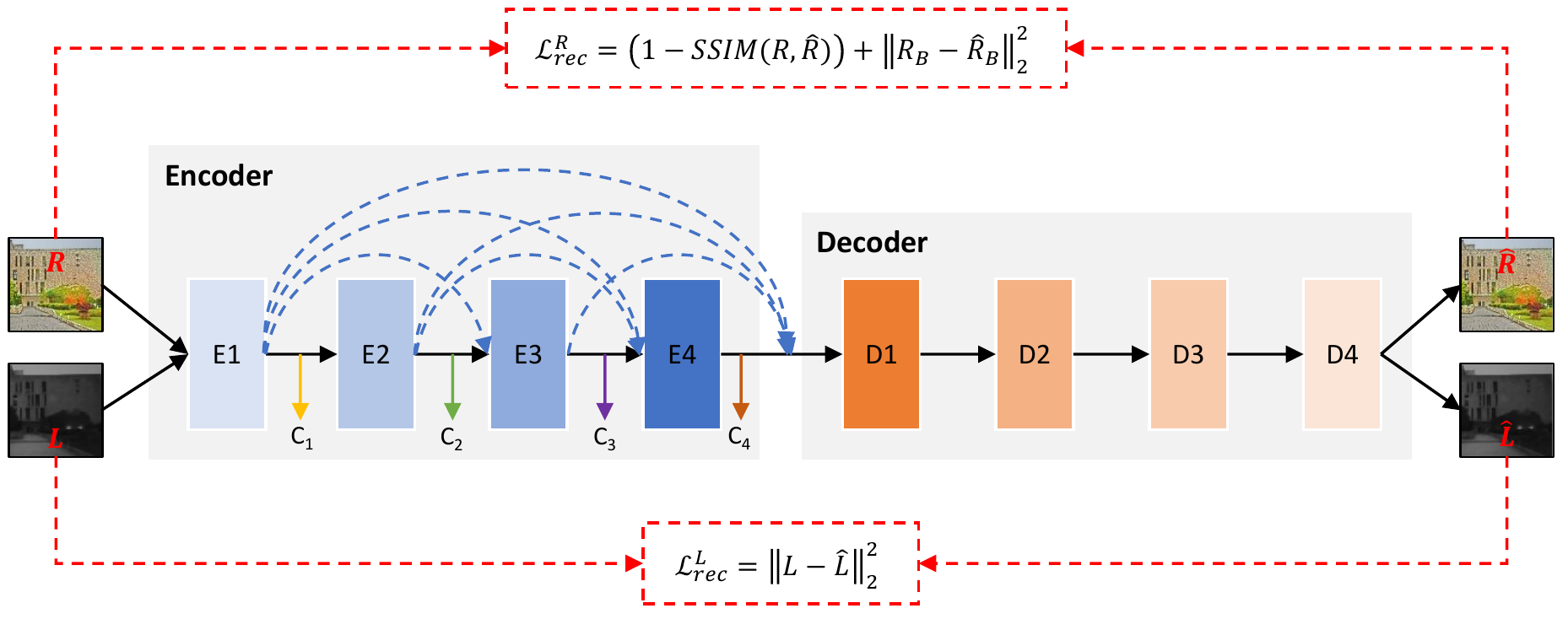}
		\caption{Detailed architecture of the self-reconstruction-based encoder-decoder for hierarchical feature learning.}
		\label{fig_featEncoder}
	\end{figure*}
	
	\subsection{Feature Encoding Module}
	Based on the reflection and illumination components generated by the image decomposition module, the next step is to build feature representations for each of these two components separately. In this work, we design a simple self-reconstruction-based encoder-decoder architecture to achieve this goal. Specifically, both the reflection and illumination components share the same feature encoding network architecture. However, these two feature encoding networks are constrained by different loss functions, i.e., tailored loss terms are designed to separately regularize the feature encoding of reflection and illumination components. 
	
	As shown in Fig. \ref{fig_featEncoder}, our proposed self-reconstruction-based encoder-decoder module involves two parts namely encoder and decoder. The encoder receives either the reflection ($R$) or illumination ($L$) component as input and progressively forms a set of hierarchical feature representations $C_1$, $C_2$, $C_3$, $C_4$. Then, the decoder takes the last-layer feature representation $C_4$ as input and progressively reconstruct the input ($\hat{R}$ or $\hat{L}$). The encoder contains four stacked 3$\times$3 convolutional layers with each convolutional layer equipped with a ReLU layer for activation. The stride of all convolutional layers is set to 1. The output feature channels of each convolutional layers in the encoder are set to 16, 32, 64, 128, respectively.

	
	Since the reflection and illumination components contain NTI degradation information in different aspects, we design tailored losses to guide the reconstruction of each component. Specifically, the losses imposed on the reflection component reconstruction include a structure loss ${\mathcal{L}}_{str}$ and a color loss ${\mathcal{L}}_{color}$, while the loss imposed on the illumination component reconstruction is a mean square error (MSE) loss ${\mathcal{L}}_{mse}$. In this way, the learned reflection feature representations will focus more on the structural and color information due to the joint guidance of ${\mathcal{L}}_{str}$ and ${\mathcal{L}}_{color}$, while the learned illumination feature representations will focus more on the luminance information. 
	
	\textit{1) Structure loss:} It is well known that the HVS is highly sensitive to image structural information and low-quality NTIs will inevitably change the structural perception \cite{wang2004SSIM}. For this, we adopt a structural similarity (SSIM) \cite{wang2004SSIM} loss between the input reflection image $R$ and its corresponding reconstructed version $\hat{R}$ for encouraging the encoder to have the capacity of extracting informative structural features. The SSIM loss is defined as follows:
	\begin{equation}
		{\mathcal{L}}_{str}=1-SSIM(R,\hat{R}),
	\end{equation}
	where $SSIM(R,\hat{R})$ computes the structural similarity score between image $R$ and image $\hat{R}$ according to the SSIM metric \cite{wang2004SSIM}.
	
	\textit{2) Color loss:} It is a common sense that NTIs will introduce color distortions and the reflection component contains almost all color information of the scene. Therefore, a simple yet effective color loss between $R$ and $\hat{R}$ is also desired, which will encourage the encoder to have the capability of extracting effective color features. Inspired by \cite{Ignatov2017}, the blurring operation can remove high frequencies of an image and promote color comparison. Thus, the following color loss is introduced:
	\begin{equation}
		{\mathcal{L}}_{color}={\left\Vert {R}_{B}-{\hat{R}}_{B} \right\Vert}_2^2,
	\end{equation}
	where ${R}_{B}$ and ${\hat{R}}_{B}$ are the blurred versions of $R$ and $\hat{R}$, respectively: 
	\begin{equation}
		{R}_{B}(i,j)=\sum_{\Omega(i,j)}{R}(i+{\Delta}_i,j+{\Delta}_j)G({\Delta}_i,{\Delta}_j),
	\end{equation}
	\begin{equation}
		{\hat{R}}_{B}(i,j)=\sum_{\Omega(i,j)}{\hat{R}}(i+{\Delta}_i,j+{\Delta}_j)G({\Delta}_i,{\Delta}_j),
	\end{equation}
	where $\Omega(i,j)$ is an image patch centered by the pixel at $(i,j)$ and $G({\Delta}_i,{\Delta}_j)$ is the 2-D Gaussian blur kernel, which can be expressed as:
	\begin{equation}
		G({\Delta}_i,{\Delta}_j)=T \cdot \text{exp}\left(-\frac{({\Delta}_i-\mu)^2+({\Delta}_j-\mu)^2}{2\sigma}\right),
	\end{equation}
	where the parameters $T=0.053$, $\mu=0$, and $\sigma=3$ are set according to \cite{Ignatov2017}.
	
	\textit{3) MSE loss:} For the reconstruction of illumination component, we only apply a simple MSE loss which is defined by the Euclidean distance between the $L$ and $\hat{L}$:
	\begin{equation}
		{\mathcal{L}}_{mse}={\left\Vert {L}-{\hat{L}} \right\Vert}_2^2.
	\end{equation}
	
	The overall loss for our feature encoding module is $\mathcal{L}_{feat}=\mathcal{L}_{str}+\mathcal{L}_{color}+\mathcal{L}_{mse}$. Constrained by these tailored losses for self-reconstruction, the content-related features and illumination-related features can be well extracted from the reflection and illumination component, respectively. 
	
	\subsection{Bilinear Pooling Module}
	We consider bilinear techniques to combine the reflection and illumination feature representations into an unified one. Bilinear models have shown powerful capability in modeling two-factor variations, such as style and content of images \cite{Tenenbaum1996SeparatingSA}, location and appearance for fine-grained recognition \cite{Lin2015BilinearCM}, temporal and spatial aspects for video analysis \cite{Simonyan2014TwoStreamCN}, etc. It also has been applied to address the BIQA problem where the synthetic and authentic distortions are modeled as the two-factor variations \cite{Zhang2020BlindIQ}. Here, we tackle the blind NTIQE problem with a similar philosophy, where the reflection-related and illumination-related degradations are modeled as the two-factor variations.
	
	Given an input NTI and its side output feature maps from the reflection and illumination encoders, $C_i^R$ and $C_i^L$ are both with the size of $h_i \times w_i \times d_i$ since the reflection and illumination encoder share the same architectures and configurations. Before performing bilinear pooling, $C_i^R$ and $C_i^L$ are separately fed into a $1 \times 1$ convolutional layer to obtain their corresponding compact version with 32 channels ($\hat{C}_i^R$ and $\hat{C}_i^L$), i.e., $h_i \times w_i \times 32$. Then, bilinear pooling is performed on $C_i^R$ and $C_i^L$ as follows:
	\begin{equation}
		B_i=(\hat{C}_i^R)^T\hat{C}_i^L,
	\end{equation}
	where the outer product $B_i$ is a vector of dimension $32 \times 32$.
	
	According to \cite{Pennec2005ARF}, bilinear representation is usually mapped from Riemannian manifold into an Euclidean space using signed square root and $\ell_2$ normalization \cite{Perronnin2010ImprovingTF}:
	\begin{equation}
		\hat{B}_i=\frac{sign({B}_i) \odot \sqrt{|{B}_i|}}{\left \Vert sign({B}_i) \odot \sqrt{|{B}_i|} \right\Vert_2^2},
	\end{equation}
	where $\odot$ means the element-wise product. Finally, the bilinear pooled feature representations over all scales are concatenated into a single vector:
	\begin{equation}
		\hat{B}=concat(\hat{B}_1,\hat{B}_2,\hat{B}_3,\hat{B}_4),
	\end{equation}
	Finally, $\hat{B}$ is fed into two fully connected layers for quality prediction, which outputs a scalar indicating the overall quality score. Here, we consider the $\ell_2$ norm as the empirical loss, which has been widely used in previous works:
	\begin{equation}
		\mathcal{L}_{quality}=\frac{1}{K}\sum_{k=1}^{K}\left \Vert Q_k-\hat{Q}_k\right\Vert_2^2,
	\end{equation}
	where $Q_k$ is the ground truth subjective quality score of the $k$-th image in a mini-batch and $\hat{Q}_k$ is the predicted quality score by DDB-Net. It is noteworthy that bilinear pooling is a global strategy and therefore our DDB-Net can receive input images with arbitrary sizes. 
	
	
	\subsection{Network Training and Testing}
	
	Our DDB-Net is trained on the target NTI quality database by minimizing the following hybrid loss function:
	\begin{equation}
		\mathcal{L}_{total}=\lambda_1\mathcal{L}_{idm}+\lambda_2\mathcal{L}_{feat}+\lambda_3\mathcal{L}_{quality}.
	\end{equation}
	where $\lambda_1$, $\lambda_2$, and $\lambda_3$ are the weights used to control the relative importance of different loss terms. The optimal weights are $\lambda_1=0.1$, $\lambda_2=0.2$, and $\lambda_3=0.7$, respectively. During training, all parameters are radomly initialized and we use the Adam optimization algorithm \cite{Kingma2015AdamAM} with a batch size of $16$. We run $100$ epoches with a learning rate of $3 \times 10^{-5}$ and use Batch normalization to stabilize the training process. All the training images are resized into $512 \times 512 \times 3$ before feeding into the network. The model is implemented by PyTorch \cite{Paszke2019PyTorchAI} with a single NVIDIA GTX 2080Ti GPU. During testing, the EAI stream will not be used. 

	\section{Experimental Results}
	
	\subsection{Experimental Setups}
	\subsubsection{Databases} For performance evaluation, we use three benchmark datasets including natural night-time image dataset (NNID) \cite{BNBT}, enhanced night-time image dataset (EHND) \footnote{https://sites.google.com/site/xiangtaooo/}, and a subset of the LIVE challenge (CLIVE) image quality dataset \cite{CLIVE}.
	
	The NNID dataset contains $2,240$ NTIs with $448$ different image contents captured by three different photographic equipments (i.e., a digital camera (Device I: Nikon D5300), a mobile phone (Device II: iPhone 8plus) and a tablet (Device III: iPad mini2)) in real-world night-time scenarios. For each image content, one device is used with five different settings to capture five images of different visual quality levels. The five settings are different for different image contents. In NNID, $1,400$ images with $280$ different image contents are captured by Nikon D5300, $640$ images with $128$ different image contents are captured by iPhone 8plus, and $200$ images with 40 different image contents are captured by iPad mini2. The images in NNID are in three different resolutions including $512 \times 512$, $1024 \times 1024$, and $2048 \times 2048$. The ground truth subjective quality score for each NTI is provided in the form of mean opinion score (MOS).
	
	The EHND dataset contains both original NTIs and their corresponding enhanced versions by different NTI enhancement algorithms. Specifically, EHND contains a total number of $1,500$ enhanced NTIs obtained by applying $15$ off-the-shelf NTI enhancement algorithms on $100$ original NTIs. Similarly, the ground truth subjective quality score, i.e., MOS, for each enhanced NTI is also provided.
	
	The CLIVE dataset contains widely diverse authentic image distortions on $1,162$ images captured using a representative variety of modern mobile devices. Each image was collected without artificially introducing any distortions beyond those occurring during capture, processing, and storage by a user's device. The MOS values obtained from the large-scale subjective studies are also available as part of this database. In the experiment, we pick out the images captured at night-time for testing. In total, we have picked out $159$ NTIs from the CLIVE dataset for testing only.
	
	\subsubsection{Protocols and Criteria} 
	We conduct experiments by following the general evaluation protocol adopted in existing learning-based BIQA studies. Specifically, we randomly divide all the images in each individual dataset into five folds with each fold contains the equal number of images. The dataset split is conducted according to source images to guarantee that there is no overlap of image content. For each time, we use four folds for training and the rest one fold for teating. The training and testing procedures are repeated five times on each database so that each image in the dataset can be tested for once. For each time, we compute four criteria to measure the model performance. The four performance criteria include Pearson linear correlation coefficient (PLCC), Spearman rank order correlation coefficient (SRCC), Kendall rank order correlation coefficient (KRCC), and root mean square error (RMSE). Among these criteria, PLCC and RMSE measure the prediction precision while SRCC and KRCC measure the prediction monotonicity. These criteria results from the five training-testing sessions are calculated respectively and averaged as the final performance.

	\begin{table*}[!t]
		\centering
		\caption{Performance Results of Different BIQA Methods on The NNID Dataset.}
		\label{tab1}
		\renewcommand\arraystretch{1.8}
		\resizebox{520pt}{!}{
			\begin{tabular}{c|cccc|cccc|cccc|cccc}
				\hline \hline
				\multicolumn{1}{c|}{\multirow{2}{*}{\textbf{Methods}}} & \multicolumn{4}{c|}{\textbf{Entire Database (2240 images)}}  & \multicolumn{4}{c|}{\textbf{Device I: Nikon D5300 (1400 images)}}  & \multicolumn{4}{c|}{\textbf{Device II: iPhone 8plus (640 images)}} & \multicolumn{4}{c}{\textbf{Device III: iPad mini2 (200 images)}} \\ \cline{2-17} 
				& \textbf{SRCC}   & \textbf{KRCC}   & \textbf{PLCC}   & \textbf{RMSE}   & \textbf{SRCC}   & \textbf{KRCC}   & \textbf{PLCC}   & \textbf{RMSE}   & \textbf{SRCC}   & \textbf{KRCC}   & \textbf{PLCC}   & \textbf{RMSE}   & \textbf{SRCC}   & \textbf{KRCC}   & \textbf{PLCC}   & \textbf{RMSE} \\ \hline \hline
				BLIINDS-II  & 0.7438 & 0.5403 & 0.7549 & 0.1119 & 0.7520 & 0.5461 & 0.7627 & 0.1108 & 0.6419 & 0.4564 & 0.6574 & 0.1103 & 0.6777 & 0.5048 & 0.7333 & 0.0892 \\ \hline
				BRISQUE   & \multicolumn{1}{c}{0.7365} & \multicolumn{1}{c}{0.5352} & \multicolumn{1}{c}{0.7452} & \multicolumn{1}{c|}{0.1132} & \multicolumn{1}{c}{0.7315} & \multicolumn{1}{c}{0.5332} & \multicolumn{1}{c}{0.7420} & \multicolumn{1}{c|}{0.1150} & \multicolumn{1}{c}{0.6445} & \multicolumn{1}{c}{0.4598} & \multicolumn{1}{c}{0.6652} & 0.1091                    & \multicolumn{1}{c}{0.5704} & \multicolumn{1}{c}{0.4166} & \multicolumn{1}{c}{0.6431} & 0.0980                    \\ \hline
				CurveletQA                                        & \multicolumn{1}{c}{0.8676} & \multicolumn{1}{c}{0.6762} & \multicolumn{1}{c}{0.8679} & \multicolumn{1}{c|}{0.0924} & \multicolumn{1}{c}{0.8937} & \multicolumn{1}{c}{0.7115} & \multicolumn{1}{c}{0.8953} & \multicolumn{1}{c|}{0.0844} & \multicolumn{1}{c}{0.8110} & \multicolumn{1}{c}{0.6147} & \multicolumn{1}{c}{0.8183} & 0.0916                    & \multicolumn{1}{c}{0.7712} & \multicolumn{1}{c}{0.5881} & \multicolumn{1}{c}{0.8217} & 0.0889                    \\ \hline
				DIIVINE                                           & \multicolumn{1}{c}{0.7744} & \multicolumn{1}{c}{0.5675} & \multicolumn{1}{c}{0.7637} & \multicolumn{1}{c|}{0.1092} & \multicolumn{1}{c}{0.7601} & \multicolumn{1}{c}{0.5545} & \multicolumn{1}{c}{0.7330} & \multicolumn{1}{c|}{0.1178} & \multicolumn{1}{c}{0.6830} & \multicolumn{1}{c}{0.4793} & \multicolumn{1}{c}{0.5844} & 0.1187                    & \multicolumn{1}{c}{0.6661} & \multicolumn{1}{c}{0.4698} & \multicolumn{1}{c}{0.6491} & 0.0998                    \\ \hline
				NRSL                                              & \multicolumn{1}{c}{0.8291} & \multicolumn{1}{c}{0.6265} & \multicolumn{1}{c}{0.8327} & \multicolumn{1}{c|}{0.0936} & \multicolumn{1}{c}{0.8165} & \multicolumn{1}{c}{0.6131} & \multicolumn{1}{c}{0.8192} & \multicolumn{1}{c|}{0.0981} & \multicolumn{1}{c}{0.7417} & \multicolumn{1}{c}{0.5417} & \multicolumn{1}{c}{0.7325} & 0.1007                    & \multicolumn{1}{c}{0.6625} & \multicolumn{1}{c}{0.4848} & \multicolumn{1}{c}{0.6903} & 0.0966                    \\ \hline
				NFERM                                             & \multicolumn{1}{c}{0.8512} & \multicolumn{1}{c}{0.6572} & \multicolumn{1}{c}{0.8556} & \multicolumn{1}{c|}{0.1099} & \multicolumn{1}{c}{0.8706} & \multicolumn{1}{c}{0.6803} & \multicolumn{1}{c}{0.8764} & \multicolumn{1}{c|}{0.1110} & \multicolumn{1}{c}{0.8122} & \multicolumn{1}{c}{0.6146} & \multicolumn{1}{c}{0.8224} & 0.1257                    & \multicolumn{1}{c}{0.7610} & \multicolumn{1}{c}{0.5727} & \multicolumn{1}{c}{0.7882} & 0.1231                    \\ \hline
				GM-LOG                                            & \multicolumn{1}{c}{0.8114} & \multicolumn{1}{c}{0.6072} & \multicolumn{1}{c}{0.8125} & 0.0985                      & \multicolumn{1}{c}{0.8135} & \multicolumn{1}{c}{0.6099} & \multicolumn{1}{c}{0.8171} & 0.0992                      & \multicolumn{1}{c}{0.7338} & \multicolumn{1}{c}{0.5338} & \multicolumn{1}{c}{0.7313} & 0.0998                    & \multicolumn{1}{c}{0.6996} & \multicolumn{1}{c}{0.5117} & \multicolumn{1}{c}{0.7107} & 0.0951                    \\ \hline
				GWH-GLBP                                          & \multicolumn{1}{c}{0.7111} & \multicolumn{1}{c}{0.5108} & \multicolumn{1}{c}{0.7098} & 0.1350                      & \multicolumn{1}{c}{0.6998} & \multicolumn{1}{c}{0.5020} & \multicolumn{1}{c}{0.6819} & 0.1382                      & \multicolumn{1}{c}{0.6383} & \multicolumn{1}{c}{0.4731} & \multicolumn{1}{c}{0.6174} & 0.1614                    & \multicolumn{1}{c}{0.6244} & \multicolumn{1}{c}{0.4547} & \multicolumn{1}{c}{0.7071} & 0.1343                    \\ \hline
				SSEQ                                              & \multicolumn{1}{c}{0.7838} & \multicolumn{1}{c}{0.5894} & \multicolumn{1}{c}{0.7865} & 0.1144                      & \multicolumn{1}{c}{0.7809} & \multicolumn{1}{c}{0.5878} & \multicolumn{1}{c}{0.7891} & 0.1258                      & \multicolumn{1}{c}{0.6735} & \multicolumn{1}{c}{0.4919} & \multicolumn{1}{c}{0.6968} & 0.1451                    & \multicolumn{1}{c}{0.6673} & \multicolumn{1}{c}{0.4617} & \multicolumn{1}{c}{0.6436} & 0.1689                    \\ \hline
				BIQME                                             & \multicolumn{1}{c}{0.8255} & \multicolumn{1}{c}{0.6185} & \multicolumn{1}{c}{0.8273} & 0.0911                      & \multicolumn{1}{c}{0.8189} & \multicolumn{1}{c}{0.6141} & \multicolumn{1}{c}{0.8245} & 0.0913                      & \multicolumn{1}{c}{0.8140} & \multicolumn{1}{c}{0.6144} & \multicolumn{1}{c}{0.8027} & 0.0972                    & \multicolumn{1}{c}{0.7935} & \multicolumn{1}{c}{0.6064} & \multicolumn{1}{c}{0.7905} & 0.1005                    \\ \hline
				ILNIQE                                            & \multicolumn{1}{c}{0.7115} & \multicolumn{1}{c}{0.5183} & \multicolumn{1}{c}{0.6335} & 0.1691                      & \multicolumn{1}{c}{0.6712} & \multicolumn{1}{c}{0.4831} & \multicolumn{1}{c}{0.6766} & 0.1679                      & \multicolumn{1}{c}{0.6949} & \multicolumn{1}{c}{0.5018} & \multicolumn{1}{c}{0.6809} & 0.1639                    & \multicolumn{1}{c}{0.7983} & \multicolumn{1}{c}{0.6086} & \multicolumn{1}{c}{0.6721} & 0.1720                    \\ \hline
				NIQE                                              & \multicolumn{1}{c}{0.5983} & \multicolumn{1}{c}{0.4220} & \multicolumn{1}{c}{0.5701} & 0.1803                      & \multicolumn{1}{c}{0.6007} & \multicolumn{1}{c}{0.4240} & \multicolumn{1}{c}{0.5859} & 0.1847                      & \multicolumn{1}{c}{0.5772} & \multicolumn{1}{c}{0.4017} & \multicolumn{1}{c}{0.5874} & 0.1811                    & \multicolumn{1}{c}{0.6591} & \multicolumn{1}{c}{0.4694} & \multicolumn{1}{c}{0.6092} & 0.1842                    \\ \hline
				BNBT                                            &
				\multicolumn{1}{c}{0.8769} & \multicolumn{1}{c}{0.6822} & \multicolumn{1}{c}{0.8784} & 0.1061                      & \multicolumn{1}{c}{0.8866} & \multicolumn{1}{c}{0.7066} & \multicolumn{1}{c}{0.8939} & 0.1020                      & \multicolumn{1}{c}{0.8632} & \multicolumn{1}{c}{0.6737} & \multicolumn{1}{c}{0.8698} & 0.1157                    & \multicolumn{1}{c}{0.8517} & \multicolumn{1}{c}{0.6890} & \multicolumn{1}{c}{0.8576} & 0.1137                    \\ \hline
				MDM                                            &
				\multicolumn{1}{c}{0.8023} & \multicolumn{1}{c}{0.6060} & \multicolumn{1}{c}{0.8039} & 0.1005                      & \multicolumn{1}{c}{0.8273} & \multicolumn{1}{c}{0.6331} & \multicolumn{1}{c}{0.7253} & 0.1323                      & \multicolumn{1}{c}{0.7741} & \multicolumn{1}{c}{0.5746} & \multicolumn{1}{c}{0.6974} & 0.1134                    & \multicolumn{1}{c}{0.7252} & \multicolumn{1}{c}{0.5542} & \multicolumn{1}{c}{0.6304} & 0.1962                    \\ \hline
				HOSA                                            &
				\multicolumn{1}{c}{0.5484} & \multicolumn{1}{c}{0.3806} & \multicolumn{1}{c}{0.5487} & 0.1416                      & \multicolumn{1}{c}{0.5547} & \multicolumn{1}{c}{0.3839} & \multicolumn{1}{c}{0.5507} & 0.1448                      & \multicolumn{1}{c}{0.4353} & \multicolumn{1}{c}{0.2983} & \multicolumn{1}{c}{0.4266} & 0.1365                    & \multicolumn{1}{c}{0.6374} & \multicolumn{1}{c}{0.4563} & \multicolumn{1}{c}{0.6202} & 0.1359                    \\ \hline \hline
				WaDIQaM                                           & \multicolumn{1}{c}{0.8272} & \multicolumn{1}{c}{0.6213} & \multicolumn{1}{c}{0.8229} & 0.0954                      & \multicolumn{1}{c}{0.8127} & \multicolumn{1}{c}{0.6258} & \multicolumn{1}{c}{0.8263} & 0.0895                      & \multicolumn{1}{c}{0.8194} & \multicolumn{1}{c}{0.6017} & \multicolumn{1}{c}{0.8101} & 0.0952                    & \multicolumn{1}{c}{0.8069} & \multicolumn{1}{c}{0.6048} & \multicolumn{1}{c}{0.8016} & 0.0937                    \\ \hline
				DBCNN                                             & \multicolumn{1}{c}{0.8938} & \multicolumn{1}{c}{0.6953} & \multicolumn{1}{c}{0.8958} & 0.0849                      & \multicolumn{1}{c}{0.8745} & \multicolumn{1}{c}{0.6779} & \multicolumn{1}{c}{0.8826} & 0.0843                      & \multicolumn{1}{c}{0.8704} & \multicolumn{1}{c}{0.6738} & \multicolumn{1}{c}{0.8796} & 0.0852                    & \multicolumn{1}{c}{0.8526} & \multicolumn{1}{c}{0.6539} & \multicolumn{1}{c}{0.8614} & 0.0893                    \\ \hline
				TSCNN                                            & \multicolumn{1}{c}{0.8618} & \multicolumn{1}{c}{0.6575} & \multicolumn{1}{c}{0.8669} & 0.0841                      & \multicolumn{1}{c}{0.8788} & \multicolumn{1}{c}{0.6863} & \multicolumn{1}{c}{0.8723} & 0.0823                      & \multicolumn{1}{c}{0.8655} & \multicolumn{1}{c}{0.6462} & \multicolumn{1}{c}{0.8638} & 0.0821                    & \multicolumn{1}{c}{0.8574} & \multicolumn{1}{c}{0.6434} & \multicolumn{1}{c}{0.8548} & 0.0851                    \\ \hline
				VCR                                             & \multicolumn{1}{c}{0.8792} & \multicolumn{1}{c}{0.6621} & \multicolumn{1}{c}{0.8744} & 0.0817                      & \multicolumn{1}{c}{0.8436} & \multicolumn{1}{c}{0.6397} & \multicolumn{1}{c}{0.8214} & 0.0896                      & \multicolumn{1}{c}{0.8812} & \multicolumn{1}{c}{0.6667} & \multicolumn{1}{c}{0.8854} & 0.0819                    & \multicolumn{1}{c}{0.8628} & \multicolumn{1}{c}{0.6533} & \multicolumn{1}{c}{0.8554} & 0.0842                    \\ \hline
				GraphBIQA                                             & \multicolumn{1}{c}{0.8618} & \multicolumn{1}{c}{0.6696} & \multicolumn{1}{c}{0.8546} & 0.1108                      & \multicolumn{1}{c}{0.8891} & \multicolumn{1}{c}{0.7034} & \multicolumn{1}{c}{0.8818} & 0.1035                      & \multicolumn{1}{c}{0.8443} & \multicolumn{1}{c}{0.6509} & \multicolumn{1}{c}{0.8425} & 0.0986                    & \multicolumn{1}{c}{0.7986} & \multicolumn{1}{c}{0.6059} & \multicolumn{1}{c}{0.7990} & 0.1074                    \\ \hline
				DDB-Net                                          & \multicolumn{1}{c}{\textbf{0.9318}} & \multicolumn{1}{c}{\textbf{0.7881}} & \multicolumn{1}{c}{\textbf{0.9311}} & \textbf{0.0745}                      & \multicolumn{1}{c}{\textbf{0.9185}} & \multicolumn{1}{c}{\textbf{0.7926}} & \multicolumn{1}{c}{\textbf{0.9203}} & \textbf{0.0762}                      & \multicolumn{1}{c}{\textbf{0.9022}} & \multicolumn{1}{c}{\textbf{0.7707}} & \multicolumn{1}{c}{\textbf{0.9003}} & \textbf{0.0801}                    & \multicolumn{1}{c}{\textbf{0.8925}} & \multicolumn{1}{c}{\textbf{0.7575}} & \multicolumn{1}{c}{\textbf{0.8918}} & \textbf{0.0824}  \\ \hline \hline
			\end{tabular}
		}
	\end{table*}

	\begin{figure*}[!t]
		\centering
		\includegraphics[width=\textwidth]{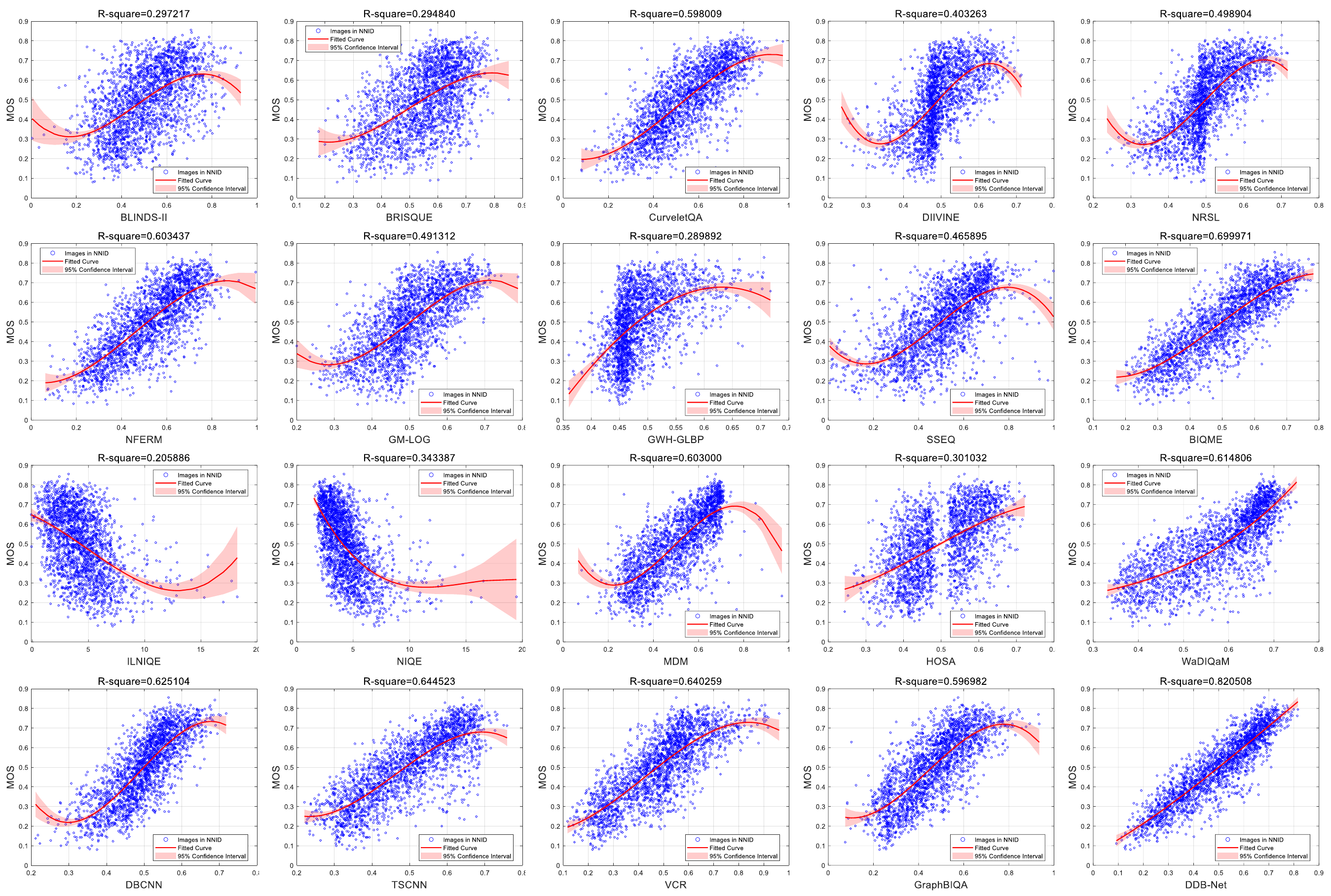}
		\caption{Scatter plots between the objective scores (predicted by BIQA methods) and subjective MOSs (provided in the NNID dataset). The scatter plots from top left to bottom right correspond to BLIINDS-II \cite{BLIINDS-II}, BRISQUE \cite{BRISQUE}, CurveletQA \cite{CurveletQA}, DIIVINE \cite{DIIVINE}, NRSL \cite{NRSL}, NFERM \cite{NFERM}, GM-LOG \cite{GM-LOG}, GWH-GLBP \cite{GWH-GLBP}, SSEQ \cite{SSEQ}, BIQME \cite{BIQME}, ILNIQE \cite{ILNIQE}, NIQE \cite{NIQE}, MDM \cite{MDM}, HOSA \cite{xu2016Blind}, WaDIQaM \cite{bosse2017deep}, DBCNN \cite{zhang2018blind}, TSCNN \cite{TSCNN}, VCR \cite{VCR}, and GraphBIQA \cite{GraphBIQA}, and the proposed DDB-Net, respectively. The corresponding fitting error in terms of the R-Square score and $95\%$ confidence interval are also provided with each plot. A higher R-Square value indicates a lower fitting error.}
		\label{fig_ScatterPlot}
	\end{figure*}
	
	\subsection{Performance Comparisons on NNID and EHND}
	Since there is always no available pristine reference for real-world NTIs, the quality evaluation of NTIs can only be performed in a no-reference manner. Therefore, we compare the performance of the proposed DDB-Net against 20 existing BIQA methods, including 15 handcrafted feature-based BIQA methods (i.e., BLIINDS-II \cite{BLIINDS-II}, BRISQUE \cite{BRISQUE}, CurveletQA \cite{CurveletQA}, DIIVINE \cite{DIIVINE}, NRSL \cite{NRSL}, NFERM \cite{NFERM}, GM-LOG \cite{GM-LOG}, GWH-GLBP \cite{GWH-GLBP}, SSEQ \cite{SSEQ}, BIQME \cite{BIQME}, ILNIQE \cite{ILNIQE}, NIQE \cite{NIQE}, BNBT \cite{BNBT}, MDM \cite{MDM}, HOSA \cite{xu2016Blind}) and five deep learning-based BIQA methods (i.e., WaDIQaM \cite{bosse2017deep}, DBCNN \cite{zhang2018blind}, TSCNN \cite{TSCNN}, VCR \cite{VCR}, and GraphBIQA \cite{GraphBIQA}). The handcrafted feature-based BIQA methods include two types: training-based and training-free. The training-based ones commonly adopt elaborately designed features to characterize the level of deviations from statistical regularities of high-quality natural images, based on which a quality prediction function is learned via supprot vector regression (SVR) \cite{Chang2011LIBSVMAL}. The training-free ones (i.e., ILNIQE \cite{ILNIQE} and NIQE \cite{NIQE}) first build a pristine statistical model from a large collection of high-quality natural images and then measure the distance between this pristine statistical model and the statistical model built on the distorted image as the estimated quality score. By contrast, the deep learning-based BIQA methods directly optimize an end-to-end function mapping from the input image to its quality score without any efforts on manual feature engineering.
	
	\begin{figure}[!t]
		\centering
		\includegraphics[width=\linewidth]{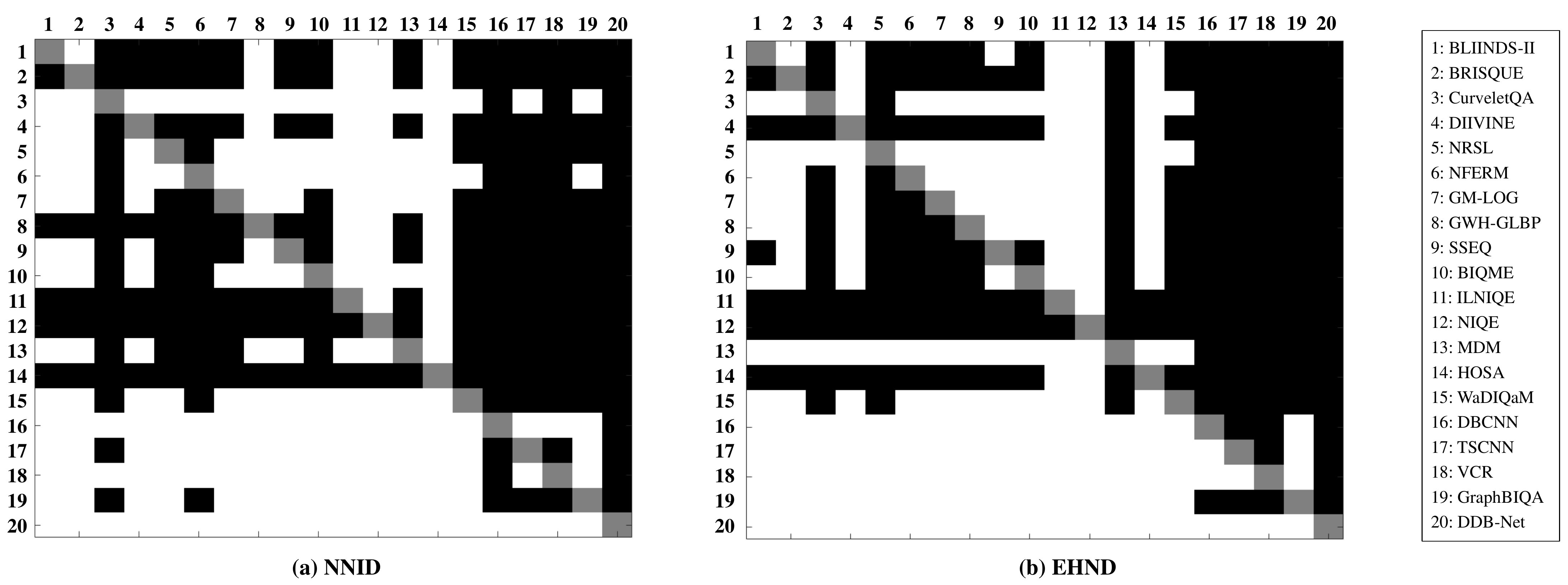}
		\caption{Significance t-test results on the NNID and EHND datasets. In the figures, a white/block small block indicates that the row model performs statistically better/worse than the column model.}
		\label{fig_SignificanceTest}
	\end{figure}
	
	\subsubsection{Comparisons on NNID} The performance comparison results of different BIQA methods on the NNID database are shown in Table \ref{tab1}. From the results, we can have the following observations. First, most training-based methods perform better than the two training-free methods (i.e., ILNIQE \cite{ILNIQE} and NIQE \cite{NIQE}) and deep learning-based methods are superior to most handcraft feature-based methods. It is reasonable because BIQA is a challenging task where training is particularly useful to model the complex non-linear relationship between the extracted features and perceived quality score, and end-to-end deep learning technique further provides an effective solution to directly establish the explicit image-to-quality mapping owing to its powerful feature representation learning capacity. Second, the existing NSS feature-based BIQA methods cannot obtain satisfactory results for evaluating NTIs because NSS is not quite suitable to characterize the degradation properties of in-the-wild NTIs. Third, the proposed DDB-Net delivers the best performance among all competitors. The reason is that we have decomposed the complex blind NTIQE task into two easier sub-tasks with each sub-task accounting for illumination perception and content perception, respectively. In such a way, the features related to the illumination perception and the content perception can be better learned to facilitate blind NTIQE.
	
	In addition to the numerical performance results, we also show the scatter plots between the objective scores (predicted by BIQA methods) and the subjective MOSs (provided in the database) in Fig. \ref{fig_ScatterPlot}. In the scatter plot, each point corresponds to an image in the NNID database. The $x$-axis represents the predicted scores by each BIQA method while the $y$-axis represents the ground truth subjective MOS. The fitted curve that characterizes the distribution of all the data points is shown in red. A good BIQA method is expected to have more compact scatter point distribution and the fitted curve should be close to the diagonal line. In addition, the corresponding fitting error in terms of the R-Square score and $95\%$ confidence interval are also provided with each plot. A higher R-Square value indicates a lower fitting error. It is obviously found that the proposed DDB-Net produces promising prediction results, which are highly consistent with subjective scores.
	
	Finally, we use a hypothesis testing approach based on t-statistics \cite{t-test} to further demonstrate whether the superiority of our proposed DDB-Net over the competitors is significant or not. In our experiment, the two-sample t-test between the pair set of PLCC values (from any two different BIQA algorithms) at the 5$\%$ significance level is conducted. Fig. \ref{fig_SignificanceTest}(a) shows the results of t-test, where the white/black small block indicates that row model performs statistically better/worse than the column model. From the results, we find that our DDB-Net always performs significantly better than all competitors, which further validates the superiority of our DDB-Net.
	

	\begin{table}[!t]
		\centering
		\caption{Performance Results of Different BIQA Methods on The EHND Dataset.}
		\label{tab2}
		\renewcommand\arraystretch{1.5}
		\resizebox{230pt}{!}{
			\begin{tabular}{c|c|c|c|c}
				\hline \hline
				\textbf{Methods} & \textbf{SRCC ($\uparrow$)}   & \textbf{KRCC ($\uparrow$)}   & \textbf{PLCC ($\uparrow$)}   & \textbf{RMSE ($\downarrow$)}   \\ \hline \hline
				BLIINDS-II   & 0.7168 & 0.5016 & 0.7026 & 0.7383 \\ \hline
				BRISQUE      & 0.7021 & 0.5077 & 0.6907 & 0.7424 \\ \hline
				CurveletQA   & 0.7525 & 0.5743 & 0.7624 & 0.6931 \\ \hline
				DIIVINE      & 0.6868 & 0.4750 & 0.6216 & 0.7593 \\ \hline
				NRSL         & 0.7853 & 0.5845 & 0.7812 & 0.6241 \\ \hline
				NFERM        & 0.7546 & 0.5683 & 0.7532 & 0.6831 \\ \hline
				GM-LOG       & 0.7915 & 0.5947 & 0.7794 & 0.6325 \\ \hline
				GWH-GLBP     & 0.7235 & 0.5074 & 0.7196 & 0.7240 \\ \hline
				SSEQ         & 0.7012 & 0.5135 & 0.6981 & 0.7383 \\ \hline
				BIQME        & 0.6916 & 0.4847 & 0.7174 & 0.7063 \\ \hline
				ILNIQE       & 0.3815 & 0.2145 & 0.4637 & 0.8034 \\ \hline
				NIQE         & 0.2723 & 0.1839 & 0.3125 & 0.8279 \\ \hline
				MDM         & 0.6959 & 0.5180 & 0.7193 & 0.7489 \\ \hline
				HOSA         & 0.3784 & 0.3179 & 0.3997 & 0.9840 \\ \hline \hline
				WaDIQaM      & 0.7528 & 0.5843 & 0.7598 & 0.6865 \\ \hline
				DBCNN        & 0.7935 & 0.6442 & 0.8051 & 0.6234 \\ \hline
				TSCNN        & 0.8174 & 0.6521 & 0.8235 & 0.6183 \\ \hline
				VCR        & 0.8213 & 0.6647 & 0.8311 & 0.6121 \\ \hline
				GraphBIQA   & 0.7967 & 0.6115 & 0.8114 & 0.6232 \\ \hline
				DDB-Net      & \textbf{0.8682} & \textbf{0.6881} & \textbf{0.8814} & \textbf{0.6032} \\ \hline \hline
			\end{tabular}
		}
	\end{table}

	\begin{figure}[!t]
		\centering
		\includegraphics[width=\linewidth]{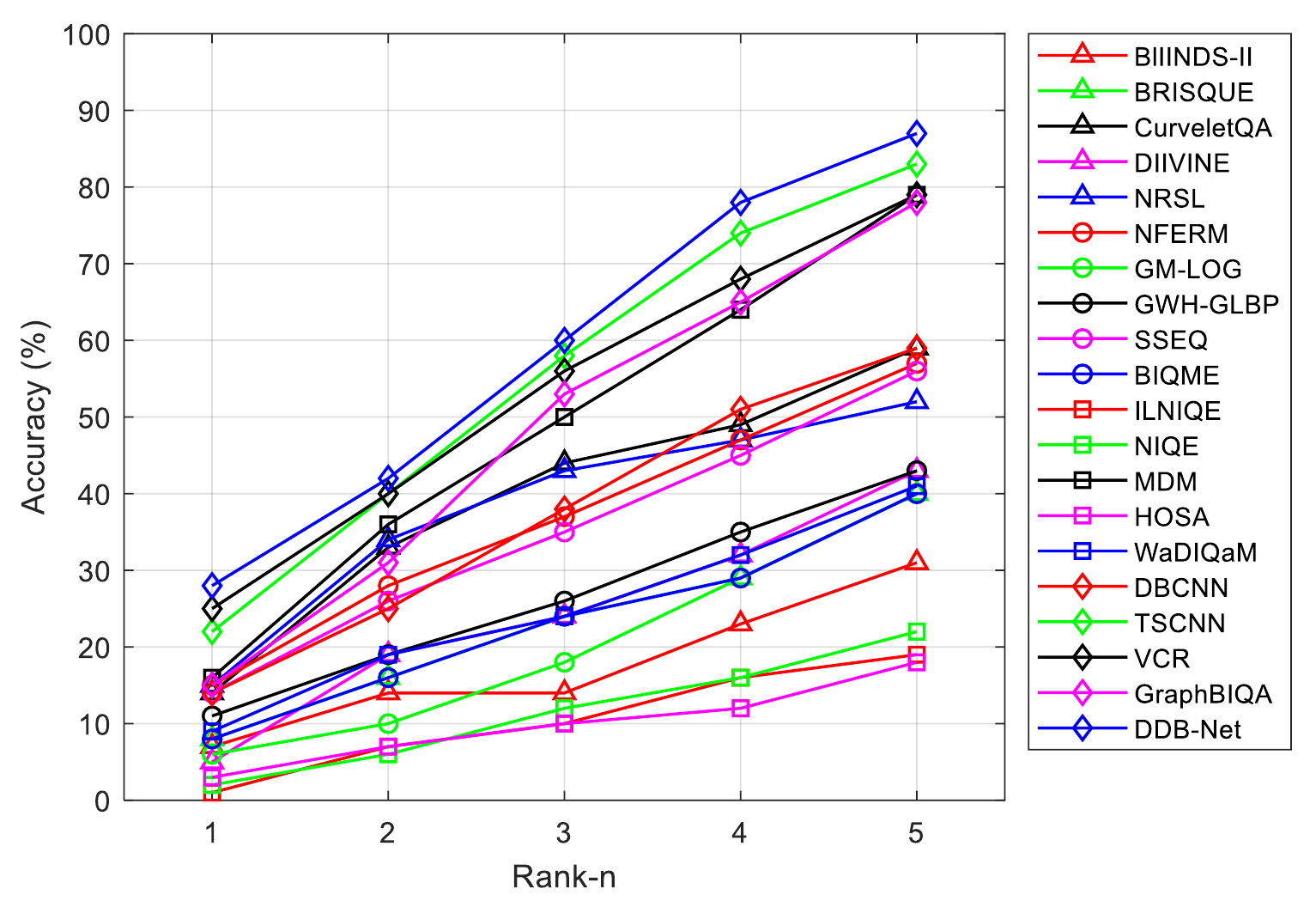}
		\caption{Comparison of the rank-1, rank-2, rank-3, rank-4, and rank-5 accuracies by different BIQA methods on the EHND dataset.}
		\label{fig_RankAccuracy}
	\end{figure}

	\subsubsection{Comparisons on EHND} 
	A well-performing NTIQE should also be able to measure the performance of different NTI quality enhancement algorithms, i.e., well evaluate different enhanced results. Actually, a certain enhancement algorithm may result in particularly bad enhanced result which may still suffer from unsatisfactory brightness and even more serious color distortions than the original raw NTI. Therefore, we also evaluate the performance of different BIQA methods on another night-time image database called EHND which contains $1,500$ images obtained by applying $15$ existing representative NTI enhancement algorithms on $100$ original NTIs. The numerical performance results of different BIQA methods on the EHND database are shown in Table \ref{tab2} and the significance t-test results are shown in Fig. \ref{fig_SignificanceTest}(b). It is observed from these results that our proposed DDB-Net again outperforms other competitors by a large margin in terms of all performance criteria. 
	
	In this case, the most important role of NTIQE is to automatically select the one with the highest visual quality from $15$ enhanced results generated from the same original NTI. Therefore, it is of great interests to conduct experiments to further compare such a kind of capability of different BIQA methods. Specifically, we measure the rank-$n$ accuracy which is closely relevant with the capability of a certain objective quality metric in selecting the optimal enhanced result from a set of candidates. Given $15$ different enhanced results associated with the same original NTI, the rank-$n$ accuracy is defined as the percentage of images whose top-$1$ result in terms of MOS appears within the top-$n$ results in terms of objective predicted score. Obviously, a higher rank-$n$ accuracy value indicates a better performance of a certain NTIQE. In Fig. \ref{fig_RankAccuracy}, we show the rank-1, rank-2, rank-3, rank-4, and rank-5 accuracy values by different BIQA methods on the EHND database. It is observed that our DDB-Net always delivers highest rank-$n$ accuracy values, indicating the best capability in selecting the one with the highest visual quality from a set of candidates.
	
	\begin{table}[!t]
		\centering
		\caption{Performance Results of Cross-Dataset Validation. Note That The MOS Values of The EHND Dataset Are Normalized into The Range of [0,1] in This Experiment.}
		\label{tab3}
		\renewcommand\arraystretch{2.0}
		\resizebox{250pt}{!}{
			\begin{tabular}{c|cccc|cccc}
				\hline \hline
				\multirow{2}{*}{\textbf{Methods}} & \multicolumn{4}{c|}{\textbf{Train on NNID \& Test on EHND}}                                               & \multicolumn{4}{c}{\textbf{Train on EHND \& Test on NNID}}                                               \\ \cline{2-9} 
				& \multicolumn{1}{c|}{\textbf{SRCC}}   & \multicolumn{1}{c|}{\textbf{KRCC}}   & \multicolumn{1}{c|}{\textbf{PLCC}}   & \textbf{RMSE}   & \multicolumn{1}{c|}{\textbf{SRCC}}   & \multicolumn{1}{c|}{\textbf{KRCC}}   & \multicolumn{1}{c|}{\textbf{PLCC}}   & \textbf{RMSE}   \\ \hline \hline
				CurveletQA               & \multicolumn{1}{c|}{0.5872} & \multicolumn{1}{c|}{0.4198} & \multicolumn{1}{c|}{0.6800} & 0.1563 & \multicolumn{1}{c|}{0.6880} & \multicolumn{1}{c|}{0.4998} & \multicolumn{1}{c|}{0.6830} & 0.1237 \\ \hline
				NRSL                     & \multicolumn{1}{c|}{0.3425} & \multicolumn{1}{c|}{0.2368} & \multicolumn{1}{c|}{0.5280} & 0.1814 & \multicolumn{1}{c|}{0.3854} & \multicolumn{1}{c|}{0.2603} & \multicolumn{1}{c|}{0.4986} & 0.1469 \\ \hline
				NFERM                    & \multicolumn{1}{c|}{0.5079} & \multicolumn{1}{c|}{0.3565} & \multicolumn{1}{c|}{0.6154} & 0.1683 & \multicolumn{1}{c|}{0.6494} & \multicolumn{1}{c|}{0.4631} & \multicolumn{1}{c|}{0.6444} & 0.1295 \\ \hline
				GM-LOG                   & \multicolumn{1}{c|}{0.4111} & \multicolumn{1}{c|}{0.2861} & \multicolumn{1}{c|}{0.5383} & 0.1793 & \multicolumn{1}{c|}{0.4063} & \multicolumn{1}{c|}{0.2758} & \multicolumn{1}{c|}{0.4710} & 0.1494 \\ \hline
				BIQME                    & \multicolumn{1}{c|}{0.5635} & \multicolumn{1}{c|}{0.4023} & \multicolumn{1}{c|}{0.6943} & 0.1545 & \multicolumn{1}{c|}{0.6000} & \multicolumn{1}{c|}{0.4170} & \multicolumn{1}{c|}{0.6472} & 0.1291 \\ \hline
				MDM                      & \multicolumn{1}{c|}{0.6521} & \multicolumn{1}{c|}{0.4698} & \multicolumn{1}{c|}{0.7190} & 0.1496 & \multicolumn{1}{c|}{0.6084} & \multicolumn{1}{c|}{0.4439} & \multicolumn{1}{c|}{0.6050} & 0.1649 \\ \hline \hline
				WaIQaM                   & \multicolumn{1}{c|}{0.6785} & \multicolumn{1}{c|}{0.4947} & \multicolumn{1}{c|}{0.7059} & 0.1599 & \multicolumn{1}{c|}{0.6923} & \multicolumn{1}{c|}{0.5182} & \multicolumn{1}{c|}{0.7037} & 0.1595 \\ \hline
				DBCNN                    & \multicolumn{1}{c|}{0.7105} & \multicolumn{1}{c|}{0.5237} & \multicolumn{1}{c|}{0.7218} & 0.1574 & \multicolumn{1}{c|}{0.7548} & \multicolumn{1}{c|}{0.5493} & \multicolumn{1}{c|}{0.7421} & 0.1362 \\ \hline
				TSCNN                    & \multicolumn{1}{c|}{0.7828} & \multicolumn{1}{c|}{0.5970} & \multicolumn{1}{c|}{0.8236} & 0.1276 & \multicolumn{1}{c|}{0.7911} & \multicolumn{1}{c|}{0.5946} & \multicolumn{1}{c|}{0.7646} & 0.1253 \\ \hline
				VCR                      & \multicolumn{1}{c|}{0.7289} & \multicolumn{1}{c|}{0.5400} & \multicolumn{1}{c|}{0.7620} & 0.1351 & \multicolumn{1}{c|}{0.8542} & \multicolumn{1}{c|}{0.6640} & \multicolumn{1}{c|}{0.8314} & 0.1157 \\ \hline
				GraphBIQA                & \multicolumn{1}{c|}{0.7233} & \multicolumn{1}{c|}{0.5384} & \multicolumn{1}{c|}{0.7758} & 0.1408 & \multicolumn{1}{c|}{0.7895} & \multicolumn{1}{c|}{0.5913} & \multicolumn{1}{c|}{0.7852} & 0.1250 \\ \hline
				DDB-Net                  & \multicolumn{1}{c|}{\textbf{0.8119}} & \multicolumn{1}{c|}{\textbf{0.6231}} & \multicolumn{1}{c|}{\textbf{0.8456}} & \textbf{0.1182} & \multicolumn{1}{c|}{\textbf{0.8691}} & \multicolumn{1}{c|}{\textbf{0.6754}} & \multicolumn{1}{c|}{\textbf{0.8572}} & \textbf{0.1116} \\ \hline \hline
			\end{tabular}
		}
	\end{table}
	
	\subsection{Cross-Dataset Validation} 
	Despite the above results have demonstrated the promising performance of our DDB-Net on each individual dataset, it remains unknown whether the model pretrained on one dataset can be well generalized to another. Therefore, we in this section conduct cross-dataset validations by training the model on one dataset and testing it on another one. Note that the ranges of MOS values of the two datasets are different, i.e., [0,1] for the NNID dataset and [0,5] for the EHND dataset, we linearly normalize the MOS values of the EHND dataset into the range of [0,1] to facilitate cross-dataset validation. The performance results of cross-dataset validations are shown in Table \ref{tab3}. In the table, we only show the results of the methods that have already achieved fairly good performance (the SRCC values are higher than 0.8) in the intra-dataset validation experiments. From Table \ref{tab3}, we can observe that the performance results obtained by training on NNID and testing on EHND are worse than the results obtained by training on EHND and testing NNID. It is expectable because the images in EHND are enhanced by different algorithms, which may suffer from extra artifacts beyond those appeared in the original raw NTIs. Despite this, our proposed DDB-Net still owns the best generalization capability among all competitors. 
	
	\subsection{Further Validation on the Night-Time Subset of CLIVE} 
\begin{figure}[!t]
		\centering
		\includegraphics[width=\linewidth]{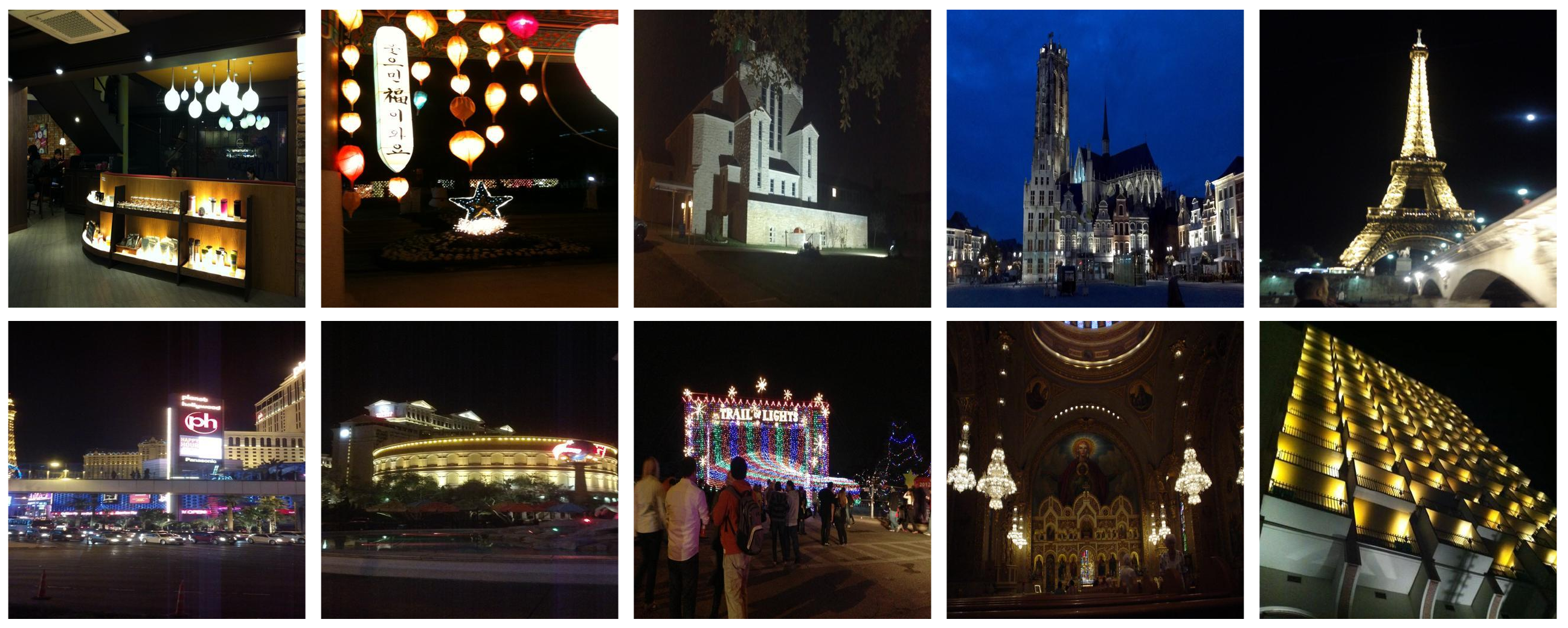}
		\caption{Examples of NTIs sampled from the CLIVE-NT subset.}
		\label{fig_CLIVE}
	\end{figure}

Besides the above two large-scale night-time image quality datasets (i.e., NNID and EHND), we are also interested in the performance of different BIQA methods on a small-scale dataset. To the best of our knowledge, except for NNID and EHND, there is no other specific dataset for NTI quality evaluation. Therefore, we pick out all the NITs from the CLIVE dataset (CLIVE-NT subset) for testing. As a result, a total number of 159 NTIs were selected to form the CLIVE-NT subset. Some examples are shown in Fig. \ref{fig_CLIVE}. Since CLIVE-NT subset is quite small, it is not suitable for training. Thus, we use the model trained on the whole NNID dataset to test the performance on the CLIVE-NT subset. The results are shown in Table. \ref{tab4}. We can observe that the performance values are consistently inferior to those on the NNID dataset. It is exceptable because the characteristics of these two datasets are quite different. Despite of this, our proposed DDB-Net is still better than other compared methods by a large margin.
	
	\begin{table}[!t]
		\centering
		\caption{Performance Results on The CLIVE-NT Subset. The Model Was Trained on The NNID Dataset and Tested on The CLIVE-NT Subset.}
		\label{tab4}
		\renewcommand\arraystretch{1.3}
		\begin{tabular}{c|c|c|c|c}
			\hline \hline
			\textbf{Methods} & \textbf{SRCC ($\uparrow$)}   & \textbf{KRCC ($\uparrow$)}   & \textbf{PLCC ($\uparrow$)}   & \textbf{RMSE ($\downarrow$)}   \\ \hline \hline
			BLIINDS-II & 0.0914 & 0.0649 & 0.1001 & 0.2117 \\ \hline
			BRISQUE    & 0.2233 & 0.1482 & 0.2288 & 0.1875 \\ \hline
			CurveletQA & 0.4240 & 0.2924 & 0.4303 & 0.1739 \\ \hline
			DIIVINE    & 0.3474 & 0.2303 & 0.3886 & 0.1775 \\ \hline
			NRSL       & 0.4470 & 0.3048 & 0.4502 & 0.1720 \\ \hline
			NFERM      & 0.3230 & 0.2190 & 0.3262 & 0.1821 \\ \hline
			GM-LOG     & 0.3681 & 0.2515 & 0.3782 & 0.1783 \\ \hline
			GWH-GLBP   & 0.2867 & 0.2005 & 0.3198 & 0.1825 \\ \hline
			SSEQ       & 0.4072 & 0.2808 & 0.4241 & 0.1745 \\ \hline
			BIQME      & 0.2589 & 0.1740 & 0.2668 & 0.1862 \\ \hline
			ILNIQE     & 0.1832 & 0.1416 & 0.1874 & 0.1946 \\ \hline
			NIQE       & 0.1619 & 0.1369 & 0.1535 & 0.1987 \\ \hline
			MDM        & 0.2525 & 0.2077 & 0.2820 & 0.1926 \\ \hline
			HOSA       & 0.2104 & 0.1437 & 0.2164 & 0.1934 \\ \hline \hline
			WaDIQaM    & 0.4537 & 0.3121 & 0.4618 & 0.1692 \\ \hline
			DBCNN      & 0.5581 & 0.3819 & 0.5462 & 0.1625 \\ \hline
			TSCNN      & 0.2323 & 0.1537 & 0.2213 & 0.1641 \\ \hline
			VCR        & 0.6024 & 0.4286 & 0.5842 & 0.1553
			\\ \hline
			GraphBIQA  & 0.4641 & 0.3285 & 0.4853 & 0.1635
			\\ \hline
			DDB-Net  & \textbf{0.6228} & \textbf{0.4593} & \textbf{0.6301} & \textbf{0.1432}
			\\ \hline \hline
		\end{tabular}
	\end{table} 
	
	\subsection{Influence of Loss Weights}
	As we have mentioned in Eq. (19), we use $\lambda_1$, $\lambda_2$, and $\lambda_3$ to control the relative importance of different loss terms. How to determine these weight values is non-trivial. Since the main task in NTIQE is to predict the quality score of an input NTI, we first set a relatively large weight to $\lambda_3$, i.e., $\lambda_3=\{0.6,0.7,0.8,0.9\}$. Then, the other two weights $\lambda_1$ and $\lambda_2$ are determined based on the constraint $\lambda_1+\lambda_2=1-\lambda_3$. We test the performance results of different weight values on the NNID dataset, as shown in Table \ref{tab5}. We can find that the best SRCC, KRCC, and RMSE are obtained when $\lambda_1=0.1$, $\lambda_2=0.2$, and $\lambda_3=0.7$ while the best PLCC is obtained when $\lambda_1=0.1$, $\lambda_2=0.1$, and $\lambda_3=0.8$. Given that the PLCC metric is related with the curve fitting process, it is less important than SRCC and KRCC. Therefore, we take $\lambda_1=0.1$, $\lambda_2=0.2$, and $\lambda_3=0.7$ as the final weights in our implementation.

	\begin{table}[!t]
		\centering
		\caption{Performance results with different weights in Eq.(19). The experiments are conducted on the NNID dataset.}
		\label{tab5}
		\renewcommand\arraystretch{1.3}
		\begin{tabular}{c|c|c|c|c|c}
			\hline \hline
			$\bf{\lambda_3}$ & $\bf{\lambda_1,\lambda_2}$ & \textbf{SRCC($\uparrow$)}   & \textbf{KRCC($\uparrow$)}   & \textbf{PLCC($\uparrow$)}   & \textbf{RMSE($\downarrow$)}   \\ \hline \hline
			\multirow{3}{*}{0.6} & 0.1, 0.3   & 0.9313 & 0.7877 & 0.9301 & 0.0751 \\ \cline{2-6} 
			& 0.2, 0.2   & 0.9314 & 0.7873 & 0.9305 & 0.0749 \\ \cline{2-6} 
			& 0.3, 0.1   & 0.9316 & 0.7876 & 0.9298 & 0.0753 \\ \hline
			\multirow{2}{*}{0.7} & 0.1, 0.2   & \textbf{0.9318} & \textbf{0.7881} & 0.9311 & \textbf{0.0745} \\ \cline{2-6} 
			& 0.2, 0.1   & 0.9313 & 0.7869 & 0.9295 & 0.0767 \\ \hline
			0.8                  & 0.1, 0.1   & 0.9315 & 0.7880 & \textbf{0.9312} & 0.0747 \\ \hline
			0.9                  & 0.05, 0.05 & 0.9310 & 0.7865 & 0.9289 & 0.0768 \\ \hline \hline
		\end{tabular}
	\end{table}

	\subsection{Computational Complexity Analysis} 
	We finally compare the computational complexity of different methods including running time, parameter number, and floating point operations (FLOPs). Note that the parameter number and FLOPs are only applicable to deep learning-based methods. The computational complexity is tested on a PC with an Intel Xeon Silver 4210 CPU @ 2.20GHz and a RTX 2080Ti GPU. The results of running time, parameter number, and FLOPs are shown in Table \ref{tab6}. We can find that deep learning-based methods generally have faster running speed than the handcrafted feature-based methods during the testing stage. Among all methods, DBCNN has the fastest running time followed by our proposed DDB-Net which ranks the second place. Among all deep learning-based methods, our DDB-Net has the lowest number of parameters and the second-lowest number of FLOPs. Note that our DDB-Net only has 0.38M parameters, which is efficient. Overall, our proposed DDB-Net can evaluate the quality of NTIs quite efficiently.
	
	\begin{table}[!t]
	\centering
	\caption{Computational Complexity Comparison of Different Methods Including Running Time, Parameter Number, and Floating Point Operations (FLOPs). Note That The Parameter Number and FLOPs Are Only Applicable to Deep Learning-Based Methods.}
	\label{tab6}
	\renewcommand\arraystretch{1.3}
	\begin{tabular}{c|c|c|c}
		\hline \hline
		\textbf{Methods} & \textbf{Time (s)} & \textbf{\# of Parameters} & \textbf{\# of FLOPs} \\ \hline \hline
		BLINDS-II  & 15.3832  & --               & --          \\ \hline
		BRISQUE    & 0.0556   & --               & --          \\ \hline
		CurveletQA & 2.4772   & --               & --          \\ \hline
		DIIVINE    & 9.5184   & --               & --          \\ \hline
		NRSL       & 0.9810   & --               & --          \\ \hline
		NFERM      & 22.5676  & --               & --          \\ \hline
		GM-LOG     & 0.3130   & --               & --          \\ \hline
		GWH-GLBP   & 0.3276   & --               & --          \\ \hline
		SSEQ       & 0.9846   & --               & --          \\ \hline
		BIQME      & 0.8317   & --               & --          \\ \hline
		ILQNIQE    & 0.3746   & --               & --          \\ \hline
		NIQE       & 0.2457   & --               & --          \\ \hline
		MDM        & 0.1344   & --               & --          \\ \hline
		HOSA       & 0.4515   & --               & --          \\ \hline \hline
		WaDIQaM    & 0.0217   & 6.29M            & 17.57G      \\ \hline
		DBCNN      & \textbf{0.0043}   & 15.31M           & 43.18G      \\ \hline
		TSCNN      & 0.0136   & \underline{1.44M}            & 13.46G      \\ \hline
		VCR        & 0.0389   & 16.66M           & 27.78G      \\ \hline
		GraphBIQA  & 0.0205   & 31.90M           & \textbf{10.76G}      \\ \hline
		DDB-Net    & \underline{0.0092}   & \textbf{0.38M}        & \underline{13.33G}      \\ \hline \hline
	\end{tabular}
	\end{table}	
	
	\section{Application: Automatic Parameter Tuning of NTI Quality Enhancement Algorithm} 
	An effective blind NTIQE should be able to well guide the optimization of NTI quality enhancement algorithms. In this section, we demonstrate this idea by applying the proposed DDB-Net to automatic parameter tuning of existing NTI quality enhancement algorithms. There are always one or several parameters in NTI quality enhancement algorithms whose optimal values vary with contents. It is challenging and time-consuming to handpick a set of parameters that work well for all image contents. A well-performing blind NTIQE is able to replace the role of humans in this task, especially when the volume of images to be processed is particularly large. 
	
\begin{figure}[!t]
		\centering
		\includegraphics[width=\linewidth]{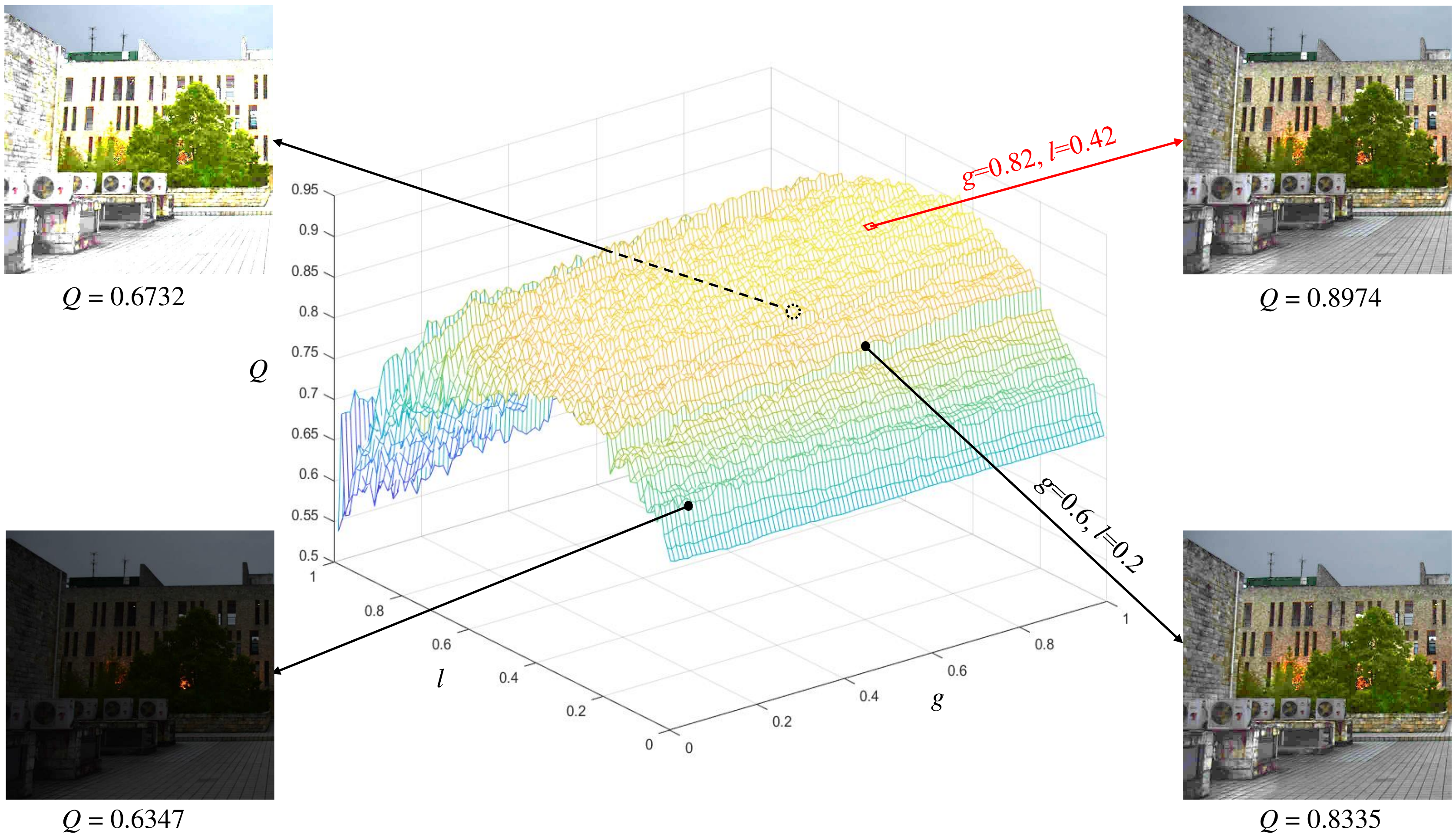}
		\caption{Automatic parameter tuning of an off-the-shelf NTI quality enhancement algorithm using the proposed DDB-Net method. Warmer color in the surface plot represents better visual quality.}
		\label{fig_ParameterTuning}
	\end{figure}

	Here, we use the LIME algorithm \cite{Guo2017LIMELI} as a representative example of NTI quality enhancement algorithm, which involves two tunable parameters $g$ and $l$. The default values are: $g=0.6$ and $l=0.2$. However, the visual quality of the final enhanced image is highly sensitive to these two parameters. Fig. \ref{fig_ParameterTuning} shows the results generated with different $g$ and $l$ values. In the figure, warmer color indicates better predicted quality of the corresponding enhanced image. The corresponding scores predicted by our DDB-Net are also shown under each image. By varying $g$ and $l$, we can obtain enhanced results with significantly different visual quality. For example, the two enhanced results in the left side of Fig. \ref{fig_ParameterTuning} still suffers from over-/under exposure problem while the two enhanced results in the right side exhibits much better visual quality with much more finer details and natural color appearance. It is found that our DDB-Net can evaluate their visual qualities consistently with human subjective perception. Furthermore, we also find that the visual quality of the upper right image is better than that of the bottom right one which is produced by using the default parameter values. It means that it is possible to adaptively determine the optimal parameter values under the guidance of our proposed DDB-Net.
	
	\section{Conclusion}
	This paper has presented a novel deep NTIQE called DDB-Net which consists of three modules namely image decomposition module, feature encoding module, and bilinear pooling module. With the help of decomposing the input NTI into two independent layer components (illumination and reflectance), the degradation features related to illumination perception and content perception are better learned and then fused with bilinear pooling to improve the performance of blind NTIQE. Experiments on two benchmark databases have demonstrated the superiority of our proposed DDB-Net. 
	
	Although our proposed DDB-Net is promising, future works towards further impoving the performance may focus on the following directions: 1) designing more efficient unsupervised solutions for image layer decomposition; 2) designing more effective loss functions to facilitate learning degradation features from each component; 3) designing more powerful feature fusion schemes by considering other variants of bilinear pooling to further improve the performance.
	
	\small
	\bibliographystyle{IEEEtran}
	\bibliography{IEEEabrv,ref}

\end{document}